\documentclass{IEEEtran}

\input{Format.tex}
\DeclarePairedDelimiter\parens{\lparen}{\rparen}
\DeclarePairedDelimiter\bracks{\lbrack}{\rbrack}
\DeclarePairedDelimiter\braces{\lbrace}{\rbrace}
\DeclarePairedDelimiter\abs{\lvert}{\rvert}

\newcommand*{\trn}{\!^{\mathsf{T}}}

\DeclareMathOperator{\C}{\mathbb{C}}

\DeclareMathOperator{\Z}{\mathbb{Z}}

\newcommand*{\E}[1]{\mathbb{E}\bracks*{#1}}

\renewcommand*{\P}[1]{\mathbb{P}\parens*{#1}}

\newcommand*{\set}[1]{\braces*{\,#1\,}}

\makeatletter
\DeclareRobustCommand\onedot{\futurelet\@let@token\@onedot}
\def\@onedot{\ifx\@let@token.\else.\null\fi\xspace}

\def\eg{\emph{e.g}\onedot} 
\def\ie{\emph{i.e}\onedot}

\def\iid{i.i.d\onedot}

\makeatother

\let\mathbf\bm

\newtcolorbox{stretchbox}[1][]{
    height fill,
    colback=white,
    colframe=black,
    #1
    }

\newtcolorbox{problem}[1]{
    breakable,
    toptitle=2.5pt,
    bottomtitle=2.5pt,
    colbacktitle=white,
    coltitle=black,
    colback=white,
    colframe=black,
    fonttitle=\large\bfseries,
    title={#1 \hfill Grade:\hspace*{0.15\paperwidth}\ }, 
}

\newtcolorbox{solution}[1]{
    breakable,
    colback=black!3,
    fonttitle=\bfseries,
    title={#1},
}

\newcommand\redout{\bgroup\markoverwith{\textcolor{red}{\rule[.5ex]{2pt}{0.4pt}}}\ULon}

\allowdisplaybreaks
\addto\captionsenglish{}

\title{ISAC MIMO Systems with OTFS Waveforms and Virtual Arrays
\thanks{The work was supported by  ARO grant W911NF2320103 and NSF grant ECCS-2320568.}
}

\author{
Kailong~Wang~\orcidlink{0000-0002-3415-0790},~\IEEEmembership{Student~Member,~IEEE},
Athina~Petropulu~\orcidlink{0000-0001-7380-7815},~\IEEEmembership{Fellow,~IEEE}%
\thanks{The authors are with the Department of Electrical and Computer Engineering, Rutgers University, Piscataway,
NJ, 08854 USA (e-mail: kailong.wang@rutgers.edu; athinap@rutgers.edu).}%
}

\begin{document}
\maketitle

\begin{abstract}
A novel Integrated Sensing-Communication (ISAC) system is proposed that can accommodate high mobility scenarios
while making efficient use of bandwidth for both communication
and sensing.
The system comprises a monostatic multiple-input
multiple-output (MIMO) radar that transmits orthogonal time
frequency space (OTFS) waveforms.
Bandwidth efficiency is
achieved by making  Doppler-delay (DD) domain bins available for shared use by the transmit antennas.
For maximum communication rate, all DD-domain bins are used as shared, but in this case, the target resolution is limited by the aperture of the receive array.
A low-complexity method is proposed for obtaining coarse estimates of the
radar targets parameters in that case.
A novel approach is also proposed to construct a virtual array (VA) for achieving a target resolution higher than that allowed by the receive array.
The VA is formed by enforcing zeros on certain time-frequency (TF) domain bins,
thereby creating private bins assigned to specific transmit antennas. The TF signals received on these private bins are orthogonal, enabling the synthesis of a VA. When combined with coarse target estimates, this approach provides high-accuracy target estimation.
To preserve DD-domain information, the introduction of private bins requires reducing the number of DD-domain symbols, resulting in a trade-off between communication rate and sensing performance.
However, even a small number of private bins is sufficient to achieve significant sensing gains with minimal communication rate loss.
The proposed system is robust to Doppler frequency shifts that arise
in high mobility scenarios.

\end{abstract}


\section{Introduction}\label{sec:intro}


\IEEEPARstart{I}{ntegrated} Sensing and Communication (ISAC) systems aim to provide both communication and sensing (C\&S) functions out of a single hardware platform.
Such systems can alleviate congestion caused by multiple sensors and transceivers, reduce device size, power consumption, and cost, and promote more efficient use of the radio spectrum.
Furthermore, the integration can significantly enhance the capabilities and performance of both communication and radar sensing functions.
As per ITU's IMT-2030 development activities~\cite{itu_imt_2030}, ISAC is one of the new usage cases in 6G systems, offering transformative capabilities in areas such as unmanned aerial vehicles (UAVs), transportation, automotive, and industrial logistics.
Whether enhancing the navigation and tracking of UAVs, robots, and vehicles or providing capabilities for collision avoidance and intrusion monitoring, ISAC can bring unprecedented advancements in safety, precision, and operational synchronization.

In general, in ISAC systems, the available resources, \eg, time, frequency, and antennas, are divided between the C\&S functions.
Here, we are interested in ISAC systems, also referred to as Dual Function Radar Communication Systems (DFRC), where both C\&S functions are performed using the same waveform as well as the same hardware platform.
Such systems offer additional benefits, maximizing spectrum efficiency, as both C\&S functions have access to all available resources.
Further, since no multiplexing of the two functions is needed, a simpler transmitter hardware is required.
These advantages are attracting significant interest in smart cities applications like intelligent vehicular networks~\cite{Kumari17vcr, daniels2018vradar,
Yuan20Bayesian,
petrov2019v2x, Liu2020radar}, and the Internet of Things (IoT)~\cite{cianca2017radiosensor, Hester2018radar,
Bletsas2018iot, Qian2018iot}.



Orthogonal frequency division multiplexing (OFDM)~\cite{Weinstein2009ofdm} waveforms and multiple-input multiple-output (MIMO)~\cite{marzetta2016fundamentals} are key technologies in the 4G and 5G systems.
For communications, OFDM is a popular approach to achieve a high communication rate and deal with frequency selective fading and has been widely used in wireless local area network (WLAN)~\cite{Nee2006ofdm802.11}
and 4G/5G mobile communications~\cite{Ergen2009Mobilebroadband,saad20206g}.
%
MIMO communication systems enable the transmission of multiple independent data streams over the same bandwidth, thus increasing the capacity of the communication channel.
In MIMO radar systems~\cite{sun2020MIMO},
the independent waveforms enable the design of flexible beams that can track multiple targets simultaneously.
Further, when the transmit waveforms are orthogonal, MIMO radar can synthesize a virtual array, which has a larger aperture than the physical receive array, thereby improving sensing performance.
ISAC MIMO systems with OFDM waveforms have been considered in~\cite{Sit18mimoofdm, bao2019precoding, xu2020dfrc,
xu2022dfrc, xu2023abandwidth,xiao2024mimoofdmisac,wu2024jcsmimoofdm}.




ISAC systems are envisioned to use high frequencies and will be deployed in high-mobility applications involving vehicular, satellite, and UAV communications.
In such scenarios, the introduced Doppler shift is significant, and the conventional time-frequency (TF) channel representation appears time-varying.
The time-varying channel destroys the orthogonality of OFDM subcarriers, resulting in Inter-Carrier Interference (ICI) and challenging OFDM signal detection.
The recently proposed Orthogonal Time Frequency Space (OTFS) modulation~\cite{Hadani2017Orthogonal} overcomes the aforementioned issues.
The OTFS approach employs Doppler-delay (DD) representation, under which the channel is sparse and appears linear and time-invariant under high Doppler.
This enables accurate equalization and signal detection in high-mobility scenarios.
 Existing works have shown that OTFS outperforms OFDM in high Doppler communication~\cite{Hadani2017Orthogonala,hong2022delay,Li2024Low,Priya2024Low}.
%
%
MIMO OTFS communication systems have attracted a lot of attention and have been studied to address the aforementioned issues in a more complex setup~\cite{Singh2022LowComplexitya,Pandey2021Low,Lu2024DelayDoppler,Srivastava2022OTFS,Qu2022Efficient}.
OTFS waveforms have also been studied for sensing in single-input single-output (SISO) systems~\cite{Raviteja2019Orthogonal,Keskin2021Radar,Zhang2023Radar} and MIMO systems~\cite{Gaudio2020Joint}.
%
%
In the aforementioned MIMO OTFS works, all antennas have unrestricted access to all DD  bins, which
 can maximize the communication rate, but  complicates the sensing at the radar receiver.
Further, such a design prevents the formation of the virtual array, limiting the sensing performance.

SISO ISAC systems using OTFS waveforms have been considered in~\cite{Zhang2021Modulation,Wu2023OTFSBased,Gaudio2019Performance,Gaudio2020Effectiveness}, demonstrating the effectiveness of OTFS as compared to OFDM.
A MIMO ISAC scenario with OTFS waveforms is considered in~\cite{Yuan2021Integrated,Shi2024Integrated},
where all DD bins are available to all transmit antennas.
In all these methods, a maximum likelihood estimation (MLE) detector with refinement is used for sensing.
The MLE detector has high complexity, which is impractical for large MIMO systems.
With the same resource arrangement,~\cite{Xia2024Achieving} proposed a generalized likelihood ratio test (GLRT) detector to reduce the complexity.
However, its sensing is still limited by the physical receive array.
MIMO ISAC OTFS systems were also proposed in~\cite{Keskin2024Integrated,Dehkordi2023BeamSpace}, with a search and track strategy.
In search mode, the DD bins are assigned to transmit antennas in an exclusive fashion to find all targets with a wide beam through the GLRT detector~\cite{Keskin2024Integrated} or Neyman-Pearson hypothesis testing~\cite{Dehkordi2023BeamSpace}.
For a single target scenario, this resource arrangement achieves orthogonality at the receiver~\cite{Keskin2024Integrated}, enabling the formation of the virtual array for improved angle resolution.
However, orthogonality is lost when multiple targets are present~\cite{Dehkordi2023BeamSpace}.
In track mode, narrow beams are used so that only one target is present in each beam pattern.
\cite{Keskin2024Integrated} jointly optimized power allocation between C\&S functions to balance performance and enable the detection of potential targets uniformly in all directions in the track mode.
\cite{Dehkordi2023BeamSpace} runs two modes asynchronously and refines sensing using an MLE detector by processing multiple blocks with varying directional beam pattern designs during the track mode.
These hybrid strategies, while leveraging the flexible beamforming capabilities of MIMO antennas, introduce complexities in transceiver design.
Besides, the search mode compromises communication rates since the resources are not wisely allocated, while the track mode radar detector may fail to detect new targets.
This paper considers the design of an ISAC system that is bandwidth-efficient and can achieve a flexible trade-off between communication and sensing functions.
In particular, the following challenges around ISAC MIMO OTFS systems are addressed.
\begin{enumerate*}[(C1)]
\item \textit{Bandwidth Allocation}:
To achieve high communication performance, which is highly desirable in next-generation systems, all transmit antennas must have access to the maximum possible bandwidth. In other words, transmit antennas should use DD bins in a shared fashion at each channel use.
At the same time, achieving a virtual array at the receive sensing array is also highly desirable as it can achieve high sensing resolution.
However, in the presence of multiple targets, OTFS waveform orthogonality is lost at the receiver due to the effect of the channel, preventing the formation of the virtual array. Further, the shared use of DD bins by antennas would result in the loss of orthogonality even in the single target case.
Exploring new mechanisms to address this challenge is crucial.
%
%
\item \textit{Low Complexity Processing}:
Although MLE is a promising approach for target estimation, its high complexity prevents its deployment in large MIMO systems.
A low-complexity approach is required for practical implementation.
\item \textit{Handling Fractional Doppler}:
In OTFS, the information symbols are multiplexed in the DD domain instead of the conventional TF domain.
As the latency and complexity of symbol detection increase with the grid size, the grid size must be kept small.
However, a limited frame length impacts Doppler resolution, making it challenging to accurately represent actual Doppler shifts using integer indices in the DD domain.
This results in fractional Doppler.
Effectively handling fractional Doppler is crucial to ensuring good communication and sensing performance.
\end{enumerate*}



\subsection{The Contribution}
To address the aforementioned challenges, we propose a novel ISAC system that comprises a fully digital monostatic MIMO radar transmitting OTFS waveforms.
The information to be transmitted is divided into blocks, which are distributed to transmit antennas and mapped to the DD domain for transmission.
Each antenna then synthesizes and transmits an OTFS waveform.
%
\subsubsection{Shared DD Bins for Coarse Estimates} We first consider the scenario in which transmit antennas place symbols on all bins (or otherwise, the DD bins are used in a \textit{shared} fashion by the transmit antennas) and propose a novel low-complexity approach for estimating target angles, velocities, and ranges.
While shared use of all DD bins results in high communication rate, the sensing performance is limited by the receive array aperture, resulting in what we refer to as \textit{coarse estimates}.
\subsubsection{Virtual Array Construction for High-Resolution Estimates} Next, we propose a virtual array construction that enables the system to achieve a resolution higher than that permitted by the physical receive array.
This is achieved by enforcing some zeros in the TF domain prior to transmission, creating a small set of \textit{private} bins uniquely associated with the transmit antennas. The data received on these private TF bins is then utilized to synthesize a virtual array.
Based on the virtual array and on discretizing the target space around the \textit{coarse estimates},  we can formulate a sparse signal recovery (SSR) problem, the solution of which provides high-resolution target estimates.
To the best of our knowledge, this is the first instance of virtual array construction in the context of ISAC MIMO OTFS with multiple targets present.
\subsubsection{Trade-Off Between Communication and Sensing}
To preserve DD-domain information, the introduction of private TF bins requires reducing the number of DD-domain symbols, resulting in a trade-off between communication rate and sensing performance. However, simulations suggest that even a small number of private bins can offer substantial sensing gains with minimal impact on the communication rate.
\subsubsection{Handling Fractional Doppler} Since the proposed virtual array leverages information in the TF domain, it achieves high sensing performance even when targets do not align with the original DD grid, \ie, in the presence of fractional Doppler. Even when fractional Doppler affects the coarse estimates, the fine-tuning provided by the SSR process effectively mitigates its impact.
\subsubsection{Cram\'er-Rao Lower Bound (CRLB) analysis}
\textcolor{black}{
We provide a Cramér-Rao Lower Bound (CRLB) analysis for the target estimates to serve as a benchmark for the quality of the obtained estimates.
}%

\subsection{Relation to the Literature}
The concept of private and shared resources, along with the use of private resources to construct a virtual array, was introduced in our earlier work~\cite{xu2023abandwidth}.
However, the waveforms in~\cite{xu2023abandwidth} were based on OFDM, with subcarriers serving as the resources.
Here, the resources are DD bins, and the virtual array construction proposed in~\cite{xu2023abandwidth} cannot be applied to the OTFS framework considered here due to the fundamentally different signal structure.
%
Relevant existing radar sensing CRLB results include \eg~\cite{Gaudio2020Effectiveness} for a SISO OTFS system, and~\cite{Gong2023Simultaneous} for a MIMO OTFS system {in the TF domain}.
We provide CLRB results for a MIMO OTFS system  {in the DD domain}.
%
Initial results of this work appeared in~\cite{wang2025dfrcotfs,wang2024virtualarraydualfunction} showing refined angle estimation beyond the limitation of physical receive arrays.
In this paper, we conduct an in-depth analysis of the proposed scheme and its advantages, including its ability to capture fractional Doppler.
We also provide a CRLB analysis of the obtained target estimates.


\section{High Communication Rate with Shared Bins}\label{sec:shared}

\begin{table}
    \centering
    \caption{List of Variable Notations}
    \label{tab:notation}
    \resizebox{\columnwidth}{!}{%
    \begin{tabular}{|l||l|}
    \hline
    $x_{n_t}[k,l]$    & Transmit symbol at antenna $n_t$ on DD bin $[k,l]$   \\ \hline
    $X_{n_t}[n,m]$    & Transmit symbol at antenna $n_t$ on TF bin $[n,m]$   \\ \hline
    $s_{n_t}(t)$      & Baseband transmit signal at antenna $n_t$            \\ \hline
    $s_{n_t}^{k,l}(t)$& Baseband transmit signal of symbol on DD bin $[k,l]$ \\ \hline
    $r_{n_r}^{k,l}(t)$& Baseband receive signal of symbol on DD bin $[k,l]$  \\ \hline
    $r_{n_r}(t)$      & Baseband receive signal at antenna $n_r$             \\ \hline
    $Y_{n_r}[n,m]$    & Receive symbol at antenna $n_r$ on TF bin $[n,m]$    \\ \hline
    $y_{n_r}[k,l]$    & Receive symbol at antenna $n_r$ on DD bin $[k,l]$    \\ \hline
    $h[k,l]$          & Effective channel on DD bin $[k,l]$                  \\ \hline
    $H[n,m]$          & Effective channel on TF bin $[n,m]$                  \\ \hline
    $h^j[k,l]$/$H^j[n,m]$    & Effective DD/TF domain channel of target/path $j$\\ \hline
    $\mathbf{x}_{n_t}$/$\mathbf{X}_{n_t}$& Vectorized DD/TF domain transmit symbols.   \\ \hline
    $\mathbf{y}_{n_r}$/$\mathbf{Y}_{n_r}$& Vectorized DD/TF domain receive symbols.   \\ \hline
    $\mathbf{h}$/$\mathbf{H}$& Vectorized DD/TF domain channel representation.   \\ \hline
    $\omega_j$        & Spatial frequency of target $j$   \\  \hline
    $A_j[k,l]$        & Angle profile of target $j$ on DD bin $[k,l]$ \\ \hline
    \end{tabular}%
    }
\end{table}

Let us consider a DFRC system comprising a monostatic MIMO radar with $N_t$ transmit antennas and $N_r$ receive antennas, transmitting OTFS waveforms to a communication receiver with $N_c$ antennas.
The carrier frequency is $f_c{\rm Hz}$, and the wavelength is $\lambda=c/f_c$ with $c$ being the speed of light.
The transmit and receive antennas form uniform linear arrays (ULA) with spacing $g_t$ and $g_r$, respectively.
The receive antennas are collocated with the transmit antennas. Thus, the radar receiver has access to the transmitted signal.
The notation used in this section is summarized in~\cref{tab:notation}.

At the transmitter, the modulated binary source data are divided into $N_t$ parallel streams, one for each transmit antenna.
Each antenna transmits packet bursts, each of a duration $T=N\Delta t$ with bandwidth $B=M\Delta f$; here $N$ is the number of subsymbols, $M$ is the number of subcarriers, $\Delta t$ is the subsymbol duration, and $\Delta f$ is the subcarrier spacing.
The orthogonality condition requires that $\Delta t \cdot \Delta f = 1$~\cite{Mohammed2021Derivation}.
In each burst, a set of $NM$ symbols are arranged on the DD grid,
\begin{align*}
    \set{\bracks*{k\Delta \nu, l\Delta\tau} \mid k=0,1,\ldots,N-1;l=0,1,\ldots,M-1},
\end{align*}
where $k$ and $l$ are Doppler and delay indices, and the grid spacing is
\(\Delta \nu={1}/{(N\Delta t)}, \  \Delta \tau ={1}/{(M\Delta f)}\).

Suppose that there are $J_T$ targets in the transmitter far-field, and let $\varphi_j\in\bracks*{-\frac{\pi}{2},\frac{\pi}{2}}$, $\nu_j=\frac{2v_j f_c}{c}$, $\tau_j=\frac{2R_j}{c}$, $\beta_j$ be the steering angle, round trip Doppler, round trip delay, and complex gain corresponding to target $j$, respectively.
We assume that the grid spacing is small enough so that $\nu_j$ and $\tau_j$ are on the grid points with integer indices, \ie,
\begin{align*}
    \nu_{j}&=k_{j}\Delta\nu, &k_{j}&\in\Z; &\tau_{j}&=l_{j}\Delta\tau, &l_{j}&\in\Z^{+}.
\end{align*}

The OTFS modulation process is illustrated in~\cref{OTFS-shared} for a $2\times 1$ MIMO system.
Let $x_{n_t}[k,l]$ be the symbol of the $n_t$-th antenna placed on DD bin $[k,l]$.
The symbols are mapped to the TF domain via the Inverse Symplectic Finite Fourier Transform (ISFFT)~\cite{Hadani2017Orthogonal}, \ie,
\begin{align}\label{eq:ISFFT}
    X_{n_t}[n,m] = \frac{1}{NM}\sum_{k=0}^{N-1}\sum_{l=0}^{M-1}x_{n_t}[k,l]e^{j2\pi\parens*{\frac{kn}{N}-\frac{ml}{M}}}.
\end{align}
The analog signal for transmission, $s(t)$, is created via the Heisenberg Transform~\cite{Hadani2017Orthogonal}, \ie,
{\small
\begin{align}\label{eq:Heisenberg}
    s_{n_t}(t) = \sum_{n=0}^{N-1}\sum_{m=0}^{M-1}X_{n_t}[n,m]g_{tx}(t-n\Delta t)e^{j2\pi m\Delta f(t-n\Delta t)},
\end{align}
}%
where $g_{tx}(t)$ is the pulse function of the transmitter.

\pgfdeclareverticalshading{rainbow}{80bp}{
    color(0bp)=(red!25); color(30bp)=(red!25); color(50bp)=(blue!25); color(80bp)=(blue!25)
    }%

\begin{figure*}
    \centering
    \resizebox{0.75\textwidth}{!}{%
    \begin{tikzpicture}[every node/.style={minimum size=.5cm-\pgflinewidth, outer sep=0pt},shading=rainbow]

    \foreach \x in {-2, 2, 6, 12, 16, 20}
    {
    \draw[dotted] (\x, 2) -- (\x, -5);
    }
    \node[above] at (0, 2) {\large DD};
    \node[above] at (4, 2) {\large TF};
    \node[above] at (9, 2) {\large Time Domain};
    \node[above] at (14, 2) {\large TF};
    \node[above] at (18, 2) {\large DD};

    \node[above left] at (-1,1) {\large TX$_1$};
    \node[above] at (0,1) {${\mathbf{x}}_{1}$};
    \foreach \x in {-0.75, -0.25, 0.25, 0.75}
    \foreach \y in {0.75, 0.25, -0.25, -0.75}
    {
    \node[fill=blue!50] at (\x, \y) {};
    }
    \draw[step=0.5cm,color=black] (-1,1) grid (1,-1);
    \draw[->] (1,1) -- (1.25,1) node[right] {\large $k$};
    \draw[->] (-1,-1) -- (-1,-1.25) node[left] {\large $l$};
    \draw[double, ->] (1,0) -- (3,0);
    \node[above] at (2,0) {\large ISFFT};

    \node[above left] at (-1,-2) {\large TX$_2$};
    \node[above] at (0,-2) {${\mathbf{x}}_{2}$};
    \foreach \x in {-0.75, -0.25, 0.25, 0.75}
    \foreach \y in {-2.25, -2.75, -3.25, -3.75}
    {
    \node[fill=red!50] at (\x, \y) {};
    }
    \draw[step=0.5cm,color=black] (-1,-2) grid (1,-4);
    \draw[->] (1,-2) -- (1.25,-2) node[right] {\large $k$};
    \draw[->] (-1,-4) -- (-1,-4.25) node[left] {\large $l$};
    \draw[double, ->] (1,-3) -- (3,-3);
    \node[above] at (2,-3) {\large ISFFT};

    \node[above] at (4,1) { ${\mathbf{X}}_{1}$};
    \fill[blue!25] (3,1) rectangle (5,-1);
    \draw[step=0.5cm,color=black] (3,1) grid (5,-1);
    \draw[->] (5,1) -- (5.25,1) node[right] {\large $n$};
    \draw[->] (3,1) -- (3,-1.25) node[left] {\large $m$};
    \draw[double, ->] (5,0) -- (7,0);
    \node[above] at (6,0) {Heisenberg};
    \node[below] at (6,0) {Transform};
    \node[right] at (7,0) {\large $s_1(t)$};
    \draw[->] (8,0) -- (8.5,0);

    \node[above] at (4,-2) {${\mathbf{X}}_{2}$};
    \fill[red!25] (3,-2) rectangle (5,-4);
    \draw[step=0.5cm,color=black] (3,-2) grid (5,-4);
    \draw[->] (5,-2) -- (5.25,-2) node[right] {\large $n$};
    \draw[->] (3,-2) -- (3,-4.25) node[left] {\large $m$};
    \draw[double, ->] (5,-3) -- (7,-3);
    \node[above] at (6,-3) {Heisenberg};
    \node[below] at (6,-3) {Transform};
    \node[right] at (7,-3) {\large $s_2(t)$};
    \draw[->] (8,-3) -- (8.5,-3);

    \draw[black] (8.5,1) rectangle (9.5,-4);
    \node[rotate=90] at (9, -1.5) {\large Channel with $J_T$ targets};
    \draw[->] (9.5,-1.5) -- (10,-1.5);
    \node[right] at (10,-1.5) {\small $r_{n_r}(t)$};
    \draw[double, ->] (11,-1.5) -- (13,-1.5);
    \node[above] at (12,-1.5) {Wigner};
    \node[below] at (12,-1.5) {Transform};

    \node[above left] at (13,-0.5) {\large RX$_{n_r}$};
    \node[above] at (14,-0.5) {$\mathbf{Y}_{n_r}$};
    \foreach \x in {14.5, 13, 13.5, 14}
    \foreach \y in {-0.5, -1, -1.5, -2}
    {
    \shade[shading angle=45] (\x,\y) rectangle +(0.5,-0.5);
    }
    \draw[step=0.5cm,color=black] (13,-0.5) grid (15,-2.5);
    \draw[->] (15,-0.5) -- (15.25,-0.5) node[right] {\large $n$};
    \draw[->] (13,-0.5) -- (13,-2.75) node[left] {\large $m$};
    \draw[double, ->] (15,-1.5) -- (17, -1.5);
    \node[above] at (16,-1.5) {\large SFFT};

    \node[above] at (18,-0.5) {\large $\mathbf{y}_{n_r}$};
    \foreach \x in {18.5, 17, 17.5, 18}
    \foreach \y in {-0.5, -1, -1.5, -2}
    {
    \shade[shading angle=45] (\x,\y) rectangle +(0.5,-0.5);
    }
    \draw[step=0.5cm,color=black] (17,-0.5) grid (19,-2.5);
    \draw[->] (19,-0.5) -- (19.25,-0.5) node[right] {\large $k$};
    \draw[->] (17,-0.5) -- (17,-2.75) node[left] {\large $l$};

    \end{tikzpicture}%
    }%
    \caption{MIMO OTFS system with all DD bins used by the transmit antennas in a shared fashion. At the radar receiver, in each DD bin, the transmitted signals and the target parameters are coupled.}%
    \label{OTFS-shared}%
\end{figure*}
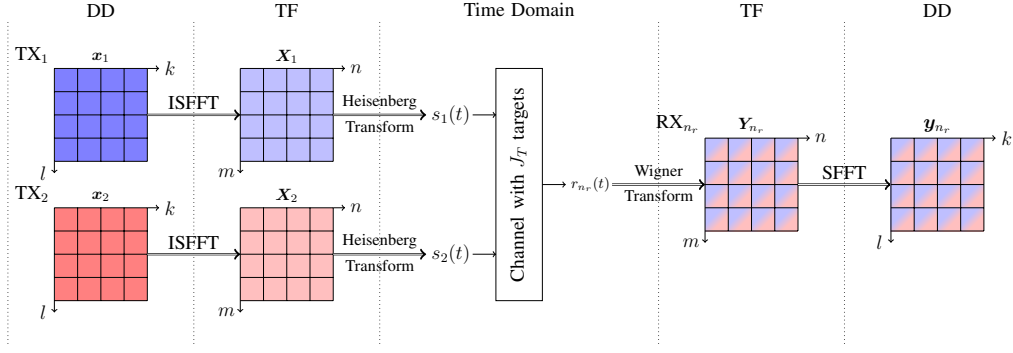%

\subsection{A Low Complexity Approach for Sensing}
The noiseless received signal at the $n_r$-th receive antenna is
\begin{align}
    r_{n_r}(t)
    &=\sum_{j=0}^{J_T-1}\sum_{n_t=0}^{N_t-1}e^{j2\pi (n_r g_r-n_t g_t)\frac{\sin(\varphi_j)}{\lambda}} \notag \\
    &\times \beta_j s_{n_t}(t-\tau_j)e^{j2\pi\nu_j(t-\tau_j)}.
\end{align}
The $n_r$-th receiver applies a matched filter $g_{rx}(t)$ and samples $r_{n_r}(t)$ over a duration $T$ at frequency $B$ (in other words, it takes the Wigner Transform~\cite{Hadani2017Orthogonal} of $r_{n_r}(t)$).
Assume $g_{tx}(t)$ and $g_{rx}(t)$ are bi-orthogonal.
Also, assume the radar detector has a fine enough angle resolution so that only one target will present in a single angular bin.
The TF domain channel input-output (I/O) relation is
\begin{align}
    Y_{n_r}[n,m]
    &=\sum_{j=0}^{J_T-1}\sum_{n_t=0}^{N_t-1}e^{j2\pi (n_rg_r-n_tg_t)\frac{\sin(\varphi_j)}{\lambda}} \notag \\
    &\times X_{n_t}[n,m]H^j[n,m], \label{eq:TFIO}
\end{align}
where
\begin{align}
    H^j[n,m]
    &=\beta_j e^{-j2\pi\nu_j\tau_j}e^{j2\pi(\nu_j n\Delta t-m\Delta f\tau_j)}. \label{eq:TFChannel}
\end{align}
The demodulated symbol of the $n_r$-th radar receive antenna corresponding to DD bin $[k,l]$ can be obtained via SFFT~\cite{Hadani2017Orthogonal},
\begin{align}
    y_{n_r}[k,l]
    &=
    \sum_{j=0}^{J_T-1}
    \sum_{n_t=0}^{N_t-1} e^{j2\pi (n_rg_r-n_tg_t)\frac{\sin(\varphi_j)}{\lambda}} \notag\\
    &\times\sum_{k'=0}^{N-1}\sum_{l'=0}^{M-1} x_{n_t}[k-k',l-l']h^j[k',l'].\label{eq:DemodulatedSymbol}
\end{align}
The $h^j [k',l']$, being the SFFT of~\cref{eq:TFChannel}, represents the DD domain sensing channel corresponding to the $j$-th target, \ie,
\begin{align}
    h^j[k', l']
    &= \frac{1}{NM}\sum_{n=0}^{N-1}\sum_{m=0}^{M-1} e^{-j2\pi\parens*{\frac{k'n}{N}-\frac{ml'}{M}}} \beta_j \notag\\
    &\times e^{-j2\pi\nu_j\tau_j}e^{j2\pi(\nu_j n\Delta t-m\Delta f\tau_j)} \label{eq:DDChannel_2}\\
    &= \frac{1}{NM}\beta_je^{-j2\pi\frac{k_j l_j}{NM}}\mathcal{G}[k',k_j]\mathcal{F}[l',l_j],  \label{eq:DDChannel_3}\\
    \mathcal{G}[k',k_j]
    &\stackrel{\Delta}{=}\sum_{n=0}^{N-1}e^{-j2\pi(k'-k_j)\frac{n}{N}} {= N\delta[k'-k_j]_N }, \label{eq:Doppler_component}\\
    \mathcal{F}[l',l_j]
    &\stackrel{\Delta}{=}\sum_{m=0}^{M-1}e^{j2\pi \frac{m}{M}(l'-l_j)} {= M\delta[l'-l_j]_M }, \label{eq:delay_component}
\end{align}
where {\([.]_N\) denotes modulo $N$ operation.}
We then get
\begin{align}
    h^j[k',l']
    &= \beta_j e^{-j2\pi\frac{k_{{j}}l_{{j}}}{NM}} \delta[k'-k_{{j}}]_N\delta[l'-l_{{j}}]_M. \label{eq:anglechannelprofile}
\end{align}
From~\cref{eq:DemodulatedSymbol,eq:anglechannelprofile}, the radar parameters $(\varphi_j, \nu_j, \tau_j)$ are coupled with the transmitted symbols in each DD bin.
While one could use an MLE approach to obtain the radar parameters, this would involve high complexity.
Next, we propose a low-complexity estimation approach suitable for practical implementation.

\subsubsection{Angle Estimation}\label{sec:shared_angle}
On lumping into $A_j[k,l]$ all terms of~\cref{eq:DemodulatedSymbol} that do not depend on $n_r$, we can rewrite~\cref{eq:DemodulatedSymbol} as
\begin{align}
    y_{n_r}[k,l]
    =\sum_{j=0}^{J_T-1}A_j[k,l]e^{j n_r\omega_j},  \label{eq:DemodulatedSymbol2} 
\end{align}
where
\begin{align}
     A_j[k,l]
     &= \sum_{n_t=0}^{N_t-1} e^{-j2\pi n_tg_t\frac{\sin(\varphi_j)}{\lambda}}\notag \\
     &\times \sum_{k'=0}^{N-1}\sum_{l'=0}^{M-1}x_{n_t}[k-k',l-l']h^j[k',l'], \label{eq:AmplitudeDecomposition} \\
    \omega_j
&=2\pi g_r{\sin(\varphi_j)}/{\lambda}. \label{angle}
\end{align}

On assuming that $N_r>J_T$, and for fixed $k$, $l$, the receive array snapshot
$\set{y_{n_r}[k,l]\mid n_r=0,1,\ldots,N_r-1}$ can be viewed as the sum of $J_T$ complex sinusoids with spatial frequencies $\omega_j$ and complex amplitudes $A_j[k,l]$.
The frequencies $\set{\omega_j \mid j=0,1,\ldots,J_T-1} $ can be estimated  
as the locations of the peaks of an $N_r$-point DFT applied to the receive array snapshot, leading to $\set{\varphi_j \mid j=0,1,\ldots,J_T-1}$ via~\cref{angle}.
The estimation can be repeated across all $NM$ bins.
This diversity improves target angle estimation.

The resolution of angle estimation, in this case, depends on the aperture of the physical receive array, \ie $(N_r - 1)g_r$.
We denote such estimates as \textit{coarse angle estimates}.

\subsubsection{Range and Velocity Estimation}\label{sec:shared_2D}
The estimation of range and velocity follows the estimation of the target angle, conducted as stated above.
That estimation
is limited by the aperture of the receive array.
If the receive array is small, there may be multiple targets corresponding to the estimated angle $\varphi_j$.
If within angle bin $\varphi_j$ there are $N_j$ targets, the complex amplitude \( A_j[.,.]\) of~\cref{eq:AmplitudeDecomposition}, corresponding to angle $\varphi_j,$ can be rewritten as
\begin{align}
    A_{\varphi_j}[k,l]
    &= \sum_{n_t=0}^{N_t-1} e^{-j2\pi n_tg_t\frac{\sin(\varphi_j)}{\lambda}}
    \notag \\
    &\times \sum_{k'=0}^{N-1}\sum_{l'=0}^{M-1}
    x_{n_t}[k-k',l-l']h^{\varphi_j}[k',l'], \label{eq:AmplitudeDecomposition_2}
\end{align}
where (based on~\cref{eq:anglechannelprofile})
{\small
    \begin{align}
    h^{\varphi_j}[k',l']
    &= \sum_{q=0}^{N_j-1}\beta_{jq} e^{-j2\pi\frac{k_{{jq}}l_{{jq}}}{NM}} \delta[k'-k_{{jq}}]_N\delta[l'-l_{{jq}}]_M, \label{eq:AngleChannel}
\end{align}
}%
$\beta_{jq}$, $k_{{jq}}$, and $l_{{jq}}$ are complex coefficient, Doppler index, and delay index of the $q$-th target with angle $\varphi_j$.
So \(h^{\varphi_j}[k,l]\) contains weighted impulses at points $\{[k_{{jq}},l_{{jq}}]\mid q=0,1,\ldots,N_j-1\}$.

Let us define $A_{\varphi_j}'[k,l]$, based on the  known transmitted symbols $x_{n_t}[k,l]$ and the already estimated (as explained above) target angles, as
\begin{align}\label{eq:AmplitudePrime}
    A_{\varphi_j}'[k,l]
    \buildrel \triangle \over = \sum_{n_t=0}^{N_t-1}e^{-j2\pi n_tg_t\frac{\sin(\varphi_j)}{\lambda}}x_{n_t}[k,l].
\end{align}
From~\cref{eq:AmplitudeDecomposition_2,eq:AngleChannel,eq:AmplitudePrime}, it can be seen that $A_{\varphi_j}[k,l]$ is a superposition of multiple weighted versions of $A_{\varphi_j}'[k,l]$ centered at points $\set{[k_{{jq}},l_{{jq}}]\mid q=0,1,\ldots,N_j-1}$,
{\small
\begin{align}
    A_{\varphi_j}[k,l]
    = \sum_{q=0}^{N_j-1}A_{\varphi_j}'[[k-k_{jq}]_N,[l-l_{jq}]_M]\beta_{jq} e^{-j2\pi\frac{k_{{jq}}l_{{jq}}}{NM}} \label{eq:corr_center}
\end{align}
}%
Therefore, the points $\set{[k_{{jq}},l_{{jq}}]\mid q=0,1,\ldots,N_j-1}$ can be found as the locations of the peaks of the 2D cross-correlation of $A_{\varphi_j}[k,l]$ and $A_{\varphi_j}'[k,l]$. Subsequently, the radar parameters can be obtained, leading to target parameters
\begin{align*}
    \nu_{jq}=k_{{jq}}\Delta \nu=\frac{2v_{jq} f_c}{c}, \quad \tau_{jq}=l_{{jq}}\Delta \tau=\frac{2R_{jq}}{c}.
\end{align*}

The ability to resolve targets in the DD domain depends on both the grid spacing and the width of the autocorrelation of $A_{\varphi_j}'[k,l]$, which can be approximated by an impulse at the origin.
The resolution and unambiguous range/velocity are
\begin{align*}
    R_{\rm res}&=\frac{c}{2M\Delta f} [\rm{m}], & R_{\rm max}&= \frac{c}{2\Delta f} [\rm{m}],  & \\
    v_{\rm res}&=\frac{\lambda}{2N\Delta t} [\rm{m/s}], & v_{\rm max}&=\frac{\lambda}{2\Delta t} [\rm{m/s}]
    . &
\end{align*}
For a large unambiguous range,  $\Delta f$ needs to be small, which results in a long symbol duration $\Delta t$.
For a large unambiguous velocity, $\Delta t$ needs to be small, resulting in a wide $\Delta f$.
In real-world systems, the trade-off between the range and the velocity resolution should be carefully considered.

\subsection{Communication}
The MIMO DD domain received signal of each pair of transmit antenna and communication antenna can be written in vector form~\cite{KollengodeRamachandran2018MIMOOTFS} as $\mathbf{y}_{n_c}=\mathbf{h}_{(n_c,n_t)}\mathbf{x}_{n_t}+\mathbf{w}_{n_c}$, where $\mathbf{y}_{n_c}$ is the {row-wise} vectorized received DD samples of the $n_c$-th communication antenna, $\mathbf{x}_{n_t}$ is the {row-wise} vectorized transmitted DD symbols from the $n_t$-th transmit antenna, $\mathbf{h}_{(n_c,n_t)}$ is the DD domain channel matrix between the $n_c$ communication antenna and the $n_t$ transmit antenna, and $\mathbf{w}_{n_c}$ is noise.
The MIMO OTFS channel I/O is then
{\small
\begin{equation}\label{eq:MIMO}
    \underbrace{
    \begin{bmatrix}
        \mathbf{y}_{1} \\
        \vdots \\
        \mathbf{y}_{N_c}
    \end{bmatrix}
    }_{\mathbf{y}}
    =
    \underbrace{
    \begin{bmatrix}
        \mathbf{h}_{(1,1)} & \cdots & \mathbf{h}_{(1,N_t)} \\
        \vdots & \ddots & \vdots \\
        \mathbf{h}_{(N_c,1)} & \cdots & \mathbf{h}_{(N_c,N_t)}
    \end{bmatrix}
    }_{\mathbf{h}}
    \underbrace{
    \begin{bmatrix}
        \mathbf{x}_{1} \\
        \vdots \\
        \mathbf{x}_{N_t}
    \end{bmatrix}
    }_{\mathbf{x}}
    +
    \underbrace{
    \begin{bmatrix}
        \mathbf{w}_{1} \\
        \vdots \\
        \mathbf{w}_{N_c}
    \end{bmatrix}
    }_{\mathbf{w}}.
\end{equation}
}%
{We assume here that each $\mathbf{h}_{(n_c, n_t)}$ is estimated using pilots~\cite{Raviteja2019Embedded,KollengodeRamachandran2018MIMOOTFS}.}
Based on the estimated channel, the transmitted symbol $\mathbf{x}$ can then be recovered via LMMSE equalization.
Since all available resources, \ie DD bins, are available to all transmit antennas in a shared fashion in the proposed system, the communication rate can be up to $N_t$ times that of the all-exclusive design.

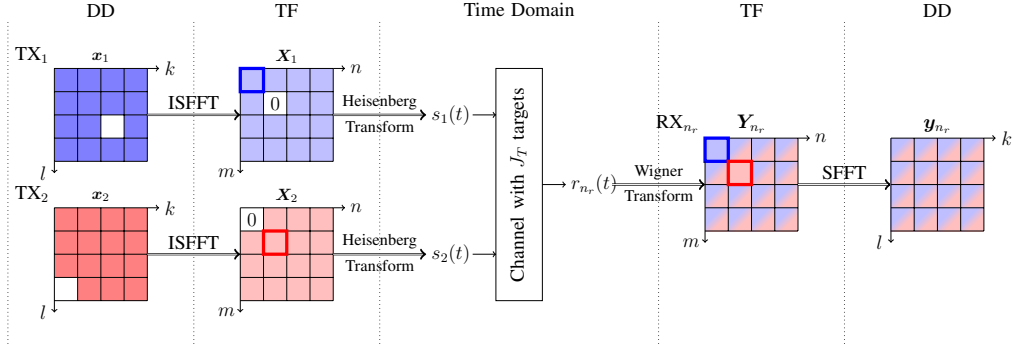
\begin{figure*}
    \centering
    \resizebox{0.75\textwidth}{!}{%
    \begin{tikzpicture}[every node/.style={minimum size=.5cm-\pgflinewidth, outer sep=0pt},shading=rainbow]

    \foreach \x in {-2, 2, 6, 12, 16, 20}
    {
    \draw[dotted] (\x, 2) -- (\x, -5);
    }
    \node[above] at (0, 2) {\large DD};
    \node[above] at (4, 2) {\large TF};
    \node[above] at (9, 2) {\large Time Domain};
    \node[above] at (14, 2) {\large TF};
    \node[above] at (18, 2) {\large DD};

    \node[above left] at (-1,1) {\large TX$_1$};
    \node[above] at (0,1) { ${\mathbf{x}}_{1}$};
    \foreach \x in {-0.75, -0.25, 0.25, 0.75}
    \foreach \y in {0.75, 0.25, -0.25, -0.75}
    {
    \node[fill=blue!50] at (\x, \y) {};
    }
    \node[fill=white] at (0.25, -0.25) {};
    \draw[step=0.5cm,color=black] (-1,1) grid (1,-1);
    \draw[->] (1,1) -- (1.25,1) node[right] {\large $k$};
    \draw[->] (-1,-1) -- (-1,-1.25) node[left] {\large $l$};
    \draw[double, ->] (1,0) -- (3,0);
    \node[above] at (2,0) {\large ISFFT};

    \node[above left] at (-1,-2) {\large TX$_2$};
    \node[above] at (0,-2) {${\mathbf{x}}_{2}$};
    \foreach \x in {-0.75, -0.25, 0.25, 0.75}
    \foreach \y in {-2.25, -2.75, -3.25, -3.75}
    {
    \node[fill=red!50] at (\x, \y) {};
    }
    \node[fill=white] at (-0.75, -3.75) {};
    \draw[step=0.5cm,color=black] (-1,-2) grid (1,-4);
    \draw[->] (1,-2) -- (1.25,-2) node[right] {\large $k$};
    \draw[->] (-1,-4) -- (-1,-4.25) node[left] {\large $l$};
    \draw[double, ->] (1,-3) -- (3,-3);
    \node[above] at (2,-3) {\large ISFFT};

    \node[above] at (4,1) { ${\mathbf{X}}_{1}$};
    \fill[blue!25] (3,1) rectangle (5,-1);
    \node[fill=white] at (3.75,0.25) {\large $0$};
    \draw[step=0.5cm,color=black] (3,1) grid (5,-1);
    \draw[->] (5,1) -- (5.25,1) node[right] {\large $n$};
    \draw[->] (3,1) -- (3,-1.25) node[left] {\large $m$};
    \draw[color=blue,line width=2pt] (3,1) rectangle (3.5,0.5);
    \draw[double, ->] (5,0) -- (7,0);
    \node[above] at (6,0) {Heisenberg};
    \node[below] at (6,0) {Transform};
    \node[right] at (7,0) {\large $s_1(t)$};
    \draw[->] (8,0) -- (8.5,0);

    \node[above] at (4,-2) {${\mathbf{X}}_{2}$};
    \fill[red!25] (3,-2) rectangle (5,-4);
    \node[fill=white] at (3.25,-2.25) {\large $0$};
    \draw[step=0.5cm,color=black] (3,-2) grid (5,-4);
    \draw[->] (5,-2) -- (5.25,-2) node[right] {\large $n$};
    \draw[->] (3,-2) -- (3,-4.25) node[left] {\large $m$};
    \draw[color=red,line width=2pt] (3.5,-2.5) rectangle (4,-3);
    \draw[double, ->] (5,-3) -- (7,-3);
    \node[above] at (6,-3) {Heisenberg};
    \node[below] at (6,-3) {Transform};
    \node[right] at (7,-3) {\large $s_2(t)$};
    \draw[->] (8,-3) -- (8.5,-3);

    \draw[black] (8.5,1) rectangle (9.5,-4);
    \node[rotate=90] at (9, -1.5) {\large Channel with $J_T$ targets};
    \draw[->] (9.5,-1.5) -- (10,-1.5);
    \node[right] at (10,-1.5) {\large $r_{n_r}(t)$};
    \draw[double, ->] (11,-1.5) -- (13,-1.5);
    \node[above] at (12,-1.5) {Wigner};
    \node[below] at (12,-1.5) {Transform};

    \node[above left] at (13,-0.5) {\large RX$_{n_r}$};
    \node[above] at (14,-0.5) {\large $\mathbf{Y}_{n_r}$};
    \foreach \x in {14.5, 13, 13.5, 14}
    \foreach \y in {-0.5, -1, -1.5, -2}
    {
    \shade[shading angle=45] (\x,\y) rectangle +(0.5,-0.5);
    }
    \fill[blue!25] (13,-0.5) rectangle (13.5,-1);
    \fill[red!25] (13.5,-1) rectangle (14,-1.5);
    \draw[step=0.5cm,color=black] (13,-0.5) grid (15,-2.5);
    \draw[color=blue,line width=2pt] (13,-0.5) rectangle (13.5,-1);
    \draw[color=red,line width=2pt] (13.5,-1) rectangle (14,-1.5);
    \draw[->] (15,-0.5) -- (15.25,-0.5) node[right] {\large $n$};
    \draw[->] (13,-1) -- (13,-2.75) node[left] {\large $m$};
    \draw[double, ->] (15,-1.5) -- (17, -1.5);
    \node[above] at (16,-1.5) {\large SFFT};

    \node[above] at (18,-0.5) {\large $\mathbf{y}_{n_r}$};
    \foreach \x in {18.5, 17, 17.5, 18}
    \foreach \y in {-0.5, -1, -1.5, -2}
    {
    \shade[shading angle=45] (\x,\y) rectangle +(0.5,-0.5);
    }
    \draw[step=0.5cm,color=black] (17,-0.5) grid (19,-2.5);
    \draw[->] (19,-0.5) -- (19.25,-0.5) node[right] {\large $k$};
    \draw[->] (17,-0.5) -- (17,-2.75) node[left] {\large $l$};

    \end{tikzpicture}%
    }%
    \caption{Each antenna zero-forces a specific time-frequency  bin, designating it as private for the other antenna. At the receiver, the signals received on these private bins are orthogonal.}
    \label{OTFS-private}
\end{figure*}

\section{Communication-Sensing Performance Trade-Off via Private Bins and Virtual Array}\label{sec:private}
When all DD bins are used by antennas in a shared fashion,
the resolution of angle estimation depends on the aperture of the physical receive array, \ie $(N_r - 1)g_r$.
Low angular resolution may cause certain target angles to be indistinguishable. Since angle estimates are used to determine target range and velocity, any errors in angle estimation will propagate to these parameters.

In this section, we propose a novel approach to surpass the resolution limits of the OTFS radar receive array
and make selective trade-offs between communication rate and sensing performance.


In our recent work~\cite{xu2023abandwidth}, we proposed a DFRC system comprising a monostatic fully digital MIMO radar transmitting OFDM waveforms.
In~\cite{xu2023abandwidth}, the OFDM subcarriers are divided into two groups: \textit{shared} and \textit{private}. On a \textit{shared} subcarrier, all antennas can transmit simultaneously, while on a \textit{private} subcarrier, only one antenna can transmit in each channel used.
When using a monostatic radar, the shared use of subcarriers results in the coupling of transmitted symbols and radar target parameters in the target echoes.
We proposed a novel, low-complexity target estimation approach to overcome the coupling and obtain target estimates based on all (shared and private) subcarriers.
Subsequently, based on those estimates and the signal received on the private subcarriers, we can formulate a virtual array that enables further improvement of the radar parameters.
Although our work in~\cite{xu2023abandwidth} provided the inspiration for the proposed work, its extension to the OTFS scenario is not trivial because the data are placed in the DD domain.
In the following, we explain how the virtual array can be formulated in the OTFS case.

\subsection{Virtual Array and Sparse Signal Recovery for Improved Target Estimation}
Let us, for simplicity, first present the idea for the case of $N_t=2$ (see~\cref{OTFS-private}).
Let $X_{p}[n,m]$ represent the TF signal of the $p$-th antenna.
Let us define TF bin $[n_p,m_p]$ to be private to antenna $p$.
In~\cref{OTFS-private}, TF bin $[0,0]$ is private to antenna $1$ (see blue square frame), and bin $[1,1]$ is private to antenna $2$ (see red square frame).
Let antenna $1$ enforce a zero in its TF bin that is private to antenna $2$, \ie set $X_1[1,1]=0$.
Similarly, let antenna $2$ set $X_{2}[0,0]=0$.
Then, let each antenna proceed to complete the OTFS modulation and transmit its signal.

The TF signal reflected by the target and received by a colocated radar receive antenna $n_r$ corresponding to private bin $[n_p,m_p]$ equals (see~\cref{eq:TFIO,eq:TFChannel})
\begin{multline}
    Y_{n_r}[n_p, m_p]
    =X_{p}[n_p,m_p] \sum_{j=0}^{J_T-1} e^{j2\pi(n_r g_r - p g_t)\frac{sin(\varphi_j)}{\lambda}} \\
    \times \beta_j e^{-j2\pi \nu_j \tau_j}e^{j2 \pi(\nu_j n_p \Delta t - m_p \Delta f \tau_j)}, \quad p=1,2. \label{16}
\end{multline}
Note that due to private nature of bin $[n_p,m_p]$, $Y_{n_r}[n_p, m_p]$ contains the signal of only one transmit antenna.
Placing the ratio
${Y_{n_r}[n_p,m_p]}/{X_{p}[n_p, m_p]}$
of all receive antennas in vector $\mathbf{r}_p\in\C^{N_r\times 1}$, we get
\begin{align}
    \mathbf{r}_p
    = \mathbf{\Phi}_r(\varphi_j, \nu_j,\tau_j ; n_p,m_p)\mathbf{1}_{J_T},
    \label{17}
\end{align}
where
$\mathbf{\Phi}_r(.,.,.; .,.)\in\C^{N_r\times J_T}$ is a matrix whose $(n_r,j)$ element
equals
\begin{align}
    e^{j2\pi(n_r g_r - p g_t)\frac{sin(\varphi_j)}{\lambda}}
    \beta_j e^{-j2\pi \nu_j \tau_j}e^{j2 \pi(\nu_j n_p \Delta t - m_p \Delta f \tau_j)}. \label{eq:SSR_ratio}
\end{align}
$\mathbf{r}_p $ can be viewed as the output of a linear array of size $N_r$.
By stacking $\mathbf{r}_1 $ and $\mathbf{r}_2 $ into vector $\mathbf{r}$, we can formulate an effective virtual array of size $2N_r$.
We can express $\mathbf{r}$ as
\begin{align}
    \mathbf{r}=[\mathbf{r}_1\trn \ \mathbf{r}_2\trn]\trn = \mathbf{\Phi}\mathbf{\beta}, \label{overcomplete}
\end{align}
where $\mathbf{\Phi}=[\mathbf{\Phi}_r\trn(\tilde \varphi_j, \tilde \nu_j,\tilde\tau_j ; n_1,m_1) \ \mathbf{\Phi}_r\trn(\tilde \varphi_j, \tilde \nu_j,\tilde\tau_j ; n_2,m_2)]\trn$ is an overcomplete matrix, with ($\tilde \varphi_j, \tilde \nu_j,\tilde\tau_j$) corresponding to a grid point of the discretized angle-Doppler-delay space.
Each element of vector $\mathbf{\beta}$ corresponds to a grid point in the target space; a non-zero element indicates the presence of a target at the corresponding grid point, and a zero value indicates the absence of a target.
Thus $\mathbf{\beta}$ will be a sparse vector.

Discretizing the entire space would lead to an unmanageable $\mathbf{\Phi}$.
Here, we leverage the target \textit{coarse estimates}, obtained as explained in~\cref{sec:shared}, and discretize the target space around those estimates to obtain a matrix $\mathbf{\Phi}$ with lower dimensions.
By restricting the target space to specific regions (\ie around the \textit{coarse estimates}), we ensure that $\mathbf{\Phi}$ remains computationally tractable while preserving the accuracy of the discretization. Further, while in general, $\mathbf{\beta}$ is not sparse, the target space around the coarse estimates will be a lot sparser than the entire target space. Therefore, we can obtain $ \mathbf{\beta} $ by solving a sparse signal recovery (SSR) problem via  orthogonal matching pursuit (OMP)~\cite{Felipe2020cs}.
%

This design can be generalized to $N_p$ private bins, with $N_p\le N_t$.
In~\cref{sec:simulation}, we will show that even a small number of $N_p$ can greatly improve the sensing resolution.

\subsubsection{Averaged SSR}

Inspired by the variance reduction in~\cref{sec:shared_angle}, \ie by averaging the angle estimates across all DD bins, we propose a modified OMP as follows.
\textcolor{black}{
The SSR estimates around the \textit{coarse estimates} exhibit low bias; therefore, the bootstrap aggregation (bagging) technique from statistical learning~\cite{hastie2009elements}
can be employed to further reduce the estimation variance.
\textcolor{black}{
We can employ $N_{\rm SSR}$ SSR solvers, where each solver uses the same discretization step, \ie $\delta\varphi$, and the same discretization width, \ie $W_{\varphi_j}$, for angle space.
In each solver, a random number $\lambda\in[0,\frac{\delta\varphi}{W_{\varphi_j}},\frac{2\delta\varphi}{W_{\varphi_j}},\ldots,1]$ is selected to construct the discretized angle space around the \textit{coarse estimates} $\varphi_j$ as $[\varphi_j-\lambda W_{\varphi_j}, \varphi_j+(1-\lambda) W_{\varphi_j}]$.
The same procedure is used to discretize the Doppler space around $\nu_j$ with $\delta\nu$ and $W_{\nu_j}$, and delay space around $\tau_j$ with $\delta\tau$ and $W_{\tau_j}$.
}%
In that way,  $N_{\rm SSR}$ matrices $\mathbf{\Phi}$ are created, corresponding to the same virtual array $\mathbf{r}$.
}%
After solving the $N_{\rm SSR}$ SSR problems
and counting the frequency of estimates from all solvers, the most frequent estimates are taken as the final refined estimates of $\mathbf{r}$.
Here we use $N_{\rm SSR}=NM$. In practice, the number of $N_{\rm SSR}$ can be determined empirically.



\subsection{Communication with Private Bins and C\&S trade-off}

While zero-forcing TF domain bins enable the formation of the virtual array and ultimately improve sensing resolution, they distort the information intended for the communication receiver.
However, it is possible to preserve the DD domain symbols despite TF zero-forcing by leaving some DD bins empty when placing symbols on the DD grid at each antenna.
To illustrate this, consider the case where each antenna leaves one DD bin empty (represented by empty squares in~\cref{OTFS-private}).
By doing so, each antenna decreases by $1$ its information-bearing symbols, which are now $NM-1$.
Each ISFFT point, $X_{p}[n,m]$, is a linear combination of all the DD domain symbols of the $p$-th antenna.
Even when antenna $p$ zero-force one TF bin to zero, the remaining $NM-1$ TF bins suffice to recover the full DD domain information; they provide $NM-1$ linear independent combinations of the antenna's $NM-1$ DD domain information-bearing symbols.

The ISFFT can be represented in vector form as
$\mathbf{X}_{p} = \mathbf{G}\mathbf{x}_{p}$
where $\mathbf{X}_{p}$ has $X_p[n,m]$ in its $(m+nM)$-th position; $\mathbf{x}_{p}$ has $x_p[k,l]$ in its $(l+kM)$-th position; and $\mathbf{G}=\mathbf{F}_N\!^{\mathsf{H}}\otimes\mathbf{F}_M$ is the ISFFT matrix with $\mathbf{F}_M$ being the $M$-points FFT matrix.
Continuing with the $N_t=2$ antenna case, if we exclude one element of $\mathbf{X}_{p}$, say the $(m_0+n_0 M)$-th element (because it is zero-forced), we lose one equation that provides a linear combination of the elements of $\mathbf{x}_{p}$.
But, if we know that $\mathbf{x}_{p}$ has a zero in a specific location, say at position $(l_0+k_0 M)$, we can still represent the TF and DD relation as
\begin{align}
 \tilde{\mathbf X}_p= \tilde{\mathbf G} \tilde{\mathbf x}_p, \label{21}
\end{align}
where $\tilde {\mathbf X}_p$ is constructed from ${\mathbf X}_{p}$ by excluding the $(m_0+n_0 M)$-th element; $\tilde{\mathbf x}_p$ is constructed from ${\mathbf x}_{p} $ by excluding the $(l_0+k_0 M)$-th element; and $\tilde {\mathbf G}$ is constructed based on ${\mathbf G}$, by removing its $(m_0+n_0M)$-th row and its $(l_0+k_0M)$-th column.
Thus, ${\mathbf x}_{p} $ can be recovered from ${\mathbf X}_{p}$.
{Let us define $\tilde{\mathbf G}^{-1}$ as the \textit{modified SFFT} associated with $(k_0,l_0), (m_0,n_0)$}.

Based on the equalized symbols, $\hat{\mathbf{x}}_{p}$, the information-bearing symbols can be obtained by taking an ISFFT to get to the TF domain, removing the zero-forced TF bins, and then applying the \textit{modified SFFT} (see \cref{21}).
This assumes knowing the locations of the empty bins in the DD domain and the zero-forced bins in the TF domain.
Notice that each private bin will reduce the information-bearing symbols by $N_t-1$, which represents the communication sensing trade-off.
Using $N_p$ private bins, the total communication rate loss is ${N_p(N_t-1)}$ as compared to the all-shared bins design of~\cref{sec:shared}.



\section{Summary of the Proposed ISAC System with VA}

The operation of the proposed system can be summarized as follows:
\begin{enumerate}
\item  Each transmit antenna, $i$, for $i=0,1,\ldots,N_t-1$, is assigned $|\mathcal{P}_i|$ private TF bins; the location of those bins is dictated by  set $\mathcal{P}_i$. The total number of private bins of the system is $N_p=\sum_{i=0}^{N_t-1}|\mathcal{P}_i|$.
\item  The $i$-th transmit antenna places symbols on the DD grid anywhere except at locations indicated by set $\mathcal{E}_i$, leaving $|\mathcal{E}_i|=N_p-|\mathcal{P}_i|$ empty bins.
Among the $N_t$ transmit antennas, a  total of $N_tNM-\sum_{i=0}^{N_t-1}(N_p-|\mathcal{P}_i|)=N_tNM-N_p(N_t-1)$ symbols are placed on the DD grid.
Let $x_i[k,l]$ denote the DD-domain symbol of antenna $i$.
\item  Each antenna converts its  DD information to the TF domain via an ISFFT. Let $X_i[n,m]$ be the TF sequence of antenna $i$.
\item  At the TF domain,  transmit antenna $i$ injects zeros at  locations $\mathcal{Z}_i$, where $|\mathcal{Z}_i|=N_p-|\mathcal{P}_i|$, After enforcing the zeros, the TF sequence becomes
\begin{equation}
    \hat{X}_i[n,m]=
        \begin{dcases*}
        X_i[n,m], &  \([n,m] \notin \mathcal{Z}_i\), \\
        0, &  \([n,m] \in \mathcal{Z}_i\).
        \end{dcases*}
\end{equation}
After the injection of zeros in the TF domain, the DD-domain equivalent, $\hat x_i[k,l]$, is the SFFT of $\hat X_i[n,m]$.
\item  Each antenna creates an OTFS symbol and transmits it.
\item  \underline{Sensing:} 
At the colocated radar receiver, each receive antenna obtains
\textit{coarse estimates} based on~\cref{eq:DemodulatedSymbol2} to~\cref{eq:corr_center}, where \(x_{i}[k,l]\) is replaced with \(\hat x_{i}[k,l]\).
Refined estimates are obtained by constructing the virtual array using the TF private bins as described in \cref{16,17,eq:SSR_ratio,overcomplete} and then solving an SSR problem as described~\cref{sec:private}.
\item  \underline{Communication}
%
At the communication receive antennas, and with knowledge of the channel, the DD symbols, $\hat{x}_{i}[k,l]$ are estimated via an LMMSE approach. To recover ${x}_{i}[k,l]$, the DD domain sequence $\hat{x}_{i}[k,l]$ is converted to a TF domain sequence via an ISFFT and then undergoes a \textit{modified SFFT} associated with the removal of rows of the SFFT matrix corresponding to the indices in \( \mathcal{Z}_i\) and columns corresponding to the indices in \(\mathcal{E}_i\) (see~\cref{21}).
\end{enumerate}

The total loss due to private bins is $N_p(N_t-1)$, while a virtual array that is  $N_p$ times larger than the physical receive array is created.

The allocation of $|\mathcal{P}_i|$ private bins and their locations ($\mathcal{P}_i$), the locations of the TF zeros ($\mathcal{Z}_i$), and the empty DD-domain bins ($\mathcal{E}_i$) can be predetermined. Their choice will be the subject of future research.
\section{Fractional Doppler}\label{sec:fractional_doppler}
In deriving the above expressions, we assumed that the targets fall on the DD grid.
In modern wideband wireless systems, the bandwidth $B$ and the number of subcarriers $M$ are sufficiently large, resulting in fine delay resolution, with the impact of fractional delay diminishing rapidly.
However, the symbol duration $T$ and the number of subsymbols $N$ are typically small to enable low-latency processing.
As a result, fractional Doppler is generally present in systems utilizing OTFS waveforms,
and parameters are related to grid points as
\begin{align*}
    \nu_{j}&=(k_{j}+\kappa_{j})\Delta\nu, &k_{j}&\in\Z, &\kappa_{j}&\in\bracks*{-0.5,0.5}.&
\end{align*}
In this section, we analyze the impact of fractional Doppler and demonstrate how the TF domain private bins and virtual array can capture fractional Doppler effects.

In the presence of fractional Doppler, the~\cref{eq:DDChannel_3} becomes
\begin{align*}
    h^j[k', l']
    &= \frac{1}{NM}\beta_je^{-j2\pi\frac{(k_j+\kappa_j)l_j}{NM}}\mathcal{G}[k',k_j+\kappa_j]\mathcal{F}[l',l_j],
\end{align*}
where
\begin{multline}
    \mathcal{G}[k',k_j+\kappa_j]
    \stackrel{\Delta}{=}\sum_{n=0}^{N-1}e^{-j2\pi(k'-(k_j+\kappa_j))\frac{n}{N}} \\
    = e^{-j2\pi(k'-(k_j+\kappa_j))\frac{N-1}{N}}\frac{sin(\pi(k'-(k_j+\kappa_j)))}{sin(\frac{\pi}{N}(k'-(k_j+\kappa_j)))}.
\end{multline}
$\mathcal{G}[k',k_j+\kappa_j]$ is a sinc-like function. When it is used in~\cref{eq:AmplitudeDecomposition_2} and then in the cross-correlation of the quantities in~\cref{eq:AmplitudeDecomposition_2,eq:AmplitudePrime}, it creates inter-Doppler-interference (IDI)~\cite{Raviteja2018Interference}, and widening of the cross-correlation peak, which  reduces the target resolution in the DD domain.

However, in our proposed approach, a coarse indication of a target can lead to a high-resolution target estimation by leveraging the virtual array and target space discretization around the coarse estimates.

\section{Cram\'er--Rao Lower Bounds of Target Estimates}
\newcommand{\bfy}{\mathbf{y}}
\newcommand{\bfh}{\mathbf{h}}
\newcommand{\bfx}{\mathbf{x}}
\newcommand{\bftt}{\mathbf{\theta}}
\newcommand{\bfSgm}{\mathbf{\Sigma}}
In this section, we derive the Cram\'er--Rao Lower Bounds (CRLB) of the unbiased target parameter estimates, providing the lower bound of the mean squared error (MSE) of the parameters.
We set $g_t=g_r=0.5\lambda$, and the spatial frequency estimated in~\cref{angle} becomes $\pi\sin(\varphi_j)$. \textcolor{black}{Initially, we assume that the targets are not on the DD grid, but in the end, in order to obtain closed-form expression, we consider the asymptotic case, where the DD grid is fine enough and the fractional Doppler and delay approach zero.}

Let us formulate the radar receive signal as
\begin{equation}
    \bfy=\mathbf{h}_s\mathbf{x}+\mathbf{w},
\end{equation}
where
\(\bfy\in\C^{N_rNM}\) is the received signal vector, formed by by stacking $NM$ row-vectorized observations of $\bfy_{n_r}$ across $N_r$ radar receive antennas; \(\bfx\in\C^{N_tNM}\) is the transmit signal vector, formed by stacking $NM$ row-vectorized DD grid of $\bfx_{n_t}$ across $N_t$ radar transmit antennas; \(\mathbf{w}\in\C^{N_rNM}\) is noise following complex Gaussian distribution $\mathcal{CN}(0,\mathbf{\Sigma})$; and \(\bfh_s\in\C^{N_rNM\times N_tNM}\) is the sensing matrix, constructed based on  $N_r\times N_t$ blocks, each representing the sensing channel between the corresponding pair of receive and transmit antennas.
Each block has dimension $NM\times NM$, and its $i$-th row is the row vectorized form of ~\cref{eq:DDChannel_3} circularly shifted to left by $i$.
The sensing channel corresponding to $y_{n_r}[k,l]$, is located in the $(n_r-1)NM+l+kM$-th row
of $\bfh_{s}$ \textcolor{black}{with non-zero value on the $(n_t-1)NM+[l-l_j]_M+[k-k_j]_N M$ columns for all $n_t\in[0,N_t-1]$ and $j\in[0,J_T-1]$}, and equals
\begin{align}
    h_{n_r,k,l}
    &= \frac{1}{NM}\sum_{j=0}^{J_T-1} e^{j\pi n_r\sin(\varphi_j)} \beta_j e^{j\phi_j}  \\
    &\times \sum_{n=0}^{N-1}e^{-j2\pi(k-\nu_j N\Delta t)\frac{n}{N}}\sum_{m=0}^{M-1}e^{j2\pi \frac{m}{M}(l-\tau_j M\Delta f)}, \notag
\\
    e^{j\phi_j} &= e^{-j2\pi\nu_j\tau_j}\sum_{n_t=0}^{N_t-1} e^{-j\pi n_t\sin(\varphi_j)}.
 \end{align}
\textcolor{black}{We should note that when the targets are not on the DD grid, \(\nu_j N\Delta t\) and \(\tau_j M\Delta f\) are not integers.}

Our parameter vector is
{\small{\begin{align}
    \bftt &= [\bftt_0\trn, \bftt_1\trn, \ldots, \bftt_{J_T-1}\trn]\trn, \label{eq:parameters} \
    \bftt_j = [\tau_j, \nu_j, \pi\sin(\varphi_j), \phi_j]\trn.
\end{align}
}}
It holds that
 $\bfy\sim\mathcal{CN}(\mathbf{h}_s\mathbf{x},\mathbf{\Sigma})$.
%
The unbiased estimate of ${\mathbf{\theta}}$, \ie $\hat{\mathbf{\theta}}$ satisfies~\cite{Poor1994intro}
\begin{align}
   \mathbf{I}^{-1}({\mathbf{\theta}})\leq\E{(\hat{\mathbf{\theta}}-{\mathbf{\theta}})(\hat{\mathbf{\theta}}-{\mathbf{\theta}})\trn},
\end{align}
where  $\mathbf{I}({\mathbf{\theta}})$ is the Fisher information matrix (FIM) and equals~\cite{key1993fundamentals}
\begin{align}\label{eq:FIM}
    \mathbf{I}(\bftt)&=\E{
    \parens*{\frac{\partial\ln\P{\bfy;\bftt}}{\partial\bftt}}
    \parens*{\frac{\partial\ln\P{\bfy;\bftt}}{\partial\bftt}}^{\mathsf{T}}} \\
    &=\E{2\Re\braces*{\parens*{\frac{\partial\bfh _s}{\partial\bftt}\bfx}^{\mathsf{H}}\bfSgm^{-1}\frac{\partial\bfh _s}{\partial\bftt}\bfx}}. \label{eq:FIM_3}
\end{align}
Let us take
$\bfSgm=N_0\mathbf{I}$.
Based on the \iid property of the information symbols, we have that
$\E{x_{n_t}[k,l]^*x_{n_t'}[k',l']}=0$,
$\forall n_t\neq n_t'$, $[k,l]\neq[k',l']$ and $\E{\abs{x_{n_t}[k,l]}^2}=P_{\rm avg}$, where $P_{\rm avg}$ denotes the average signal power.
Then, we have that
\begin{align}
    \mathbf{I}(\bftt)
    &=\frac{2 P_{\rm avg}}{N_0}\Re\braces*{\parens*{\frac{\partial\bfh_s}{\partial\bftt}}^{\mathsf{H}}\frac{\partial\bfh_s}{\partial\bftt}}
    \label{eq:FIM_5}
\end{align}
which is a block diagonal matrix with the $(i,j)$ block equal to
\begin{align}
    \mathbf{I}_{\bftt_j\bftt_j}
    &= \frac{2 P_{\rm avg}}{N_0}\Re\braces*{\parens*{\frac{\partial\bfh_s}{\partial\bftt_j}}^{\mathsf{H}}\frac{\partial\bfh_s}{\partial\bftt_j}}. \label{eq:FIM_6}
\end{align}
Since~\cref{eq:FIM_5} is a diagonal matrix, the CRLB $\mathbf{I}^{-1}(\bftt)$ of all parameters can be obtained by the inversion of FIM of each target as $\mathbf{I}_{\bftt_j\bftt_j}^{-1}$.
The CRLBs of the target $j$, which are the diagonal elements of $\mathbf{I}_{\bftt_j\bftt_j}^{-1}$ is obtained as (see~\cref{apd:crlb} for details)
\begin{align}
    {\tau_j}^{\rm CRLB}
    &= \frac{N_0}{2\abs{\beta_j}^2 P_{\rm avg}}\frac{1}{N_r}\cdot\frac{3}{\pi^2 (\Delta f)^2 (M^2-1)}, \label{eq:crlb_tau}\\
    {\nu_j}^{\rm CRLB}
    &= \frac{N_0}{2\abs{\beta_j}^2 P_{\rm avg}}\frac{1}{N_r}\cdot\frac{3}{\pi^2 (\Delta t)^2 (N^2-1)}, \label{eq:crlb_nu}\\
    {\pi\sin(\varphi_j)}^{\rm CRLB}
    &= \frac{N_0}{2\abs{\beta_j}^2 P_{\rm avg}}\frac{1}{N_r}\cdot\frac{12\cdot7}{(N_r-1)(7N_r-1)} \label{eq:crlb_angle}
\end{align}
Assuming the complex coefficients have unit magnitude $\abs{\beta_j}^2=1$ and the information symbol constellation has unit average power $P_{\rm avg}=1$, the scalar term in the CRLBs becomes $\frac{N_0}{2\abs{\beta_j}^2 P_{\rm avg}}=\frac{1}{2{\rm SNR}}$.

\section{Simulation Results}\label{sec:simulation}
\begin{table}
    \centering
    \caption{System Parameters }
    \label{tab:sys_parameters}
    \begin{tabular}{|c|c|c|}
    \hline
    Symbol      & Parameter             & Value         \\ \hline \hline
    \(M\)        & Number of subcarriers & $128$         \\ \hline
    \(N\)        & Number of subsymbols  & $64$          \\ \hline
    \(\Delta f\) & Subcarrier spacing    & $120$ KHz  \\ \hline
    \(f_c\)      & Carrier frequency     & $24.25$ GHz      \\ \hline
    \(g_t\)      & Transmit antenna spacing & $0.5\lambda$ \\ \hline
    \(g_r\)      & Receive antenna spacing  & $0.5\lambda$ \\ \hline
    \end{tabular}%
\end{table}

\begin{table}
    \centering
    \caption{Channel Parameters
    }
    \label{tab:share_parameters}
    \begin{tabular}{|c|c|c|}
    \hline
    Symbol      & Parameter             & Value         \\ \hline \hline
    \(J_T/J\)        & Number of targets/paths     & $3$           \\ \hline
    \(\varphi_j\) & Angle of targets/paths \  & $[7, -14, 22]^{\circ}$ \\ \hline
    \(R_j\)      & Range of targets/paths   & $[73.48, 64.29, 45.92]{\rm{m}}$ \\ \hline
    \(v_j\)      & Velocity of targets/paths  & $[54.54, -98.17, 76.36]{\rm{m/s}}$ \\ \hline
    \end{tabular}%
\end{table}


In this section, we present simulation results to demonstrate the performance of the proposed system.
The default system parameters follow the 5G NR high-frequency standard~\cite{vook20185gnr} and are shown in~\cref{tab:sys_parameters}.
The high-Doppler channel is simulated with high target speed, $v_j\approx100$m/s, to align with the requirements outlined in ITU-R M.2410-0~\cite{itu2020minrequirements} and is primarily intended for high-speed trains.
Both the sensing and communication channels consist of three targets/paths, with each pair of transmit and receive/communication antennas exhibiting distinct unit-magnitude complex path gains.
The default information symbols are QPSK, and the default signal-to-noise ratio (SNR) is $20$dB.

We also evaluate the applicability of OFDM in this ISAC MIMO scenario and compare the results of the method proposed in~\cref{sec:shared} with the coarse estimates from~\cite{xu2023abandwidth}, which correspond to a MIMO OFDM system.

\subsection{Shared Use of DD Bins to Maximize Communication Rate}



\begin{figure}
    \centering
    \begin{subfigure}[t]{0.45\columnwidth}
        \centering
        \includegraphics[width=\columnwidth]{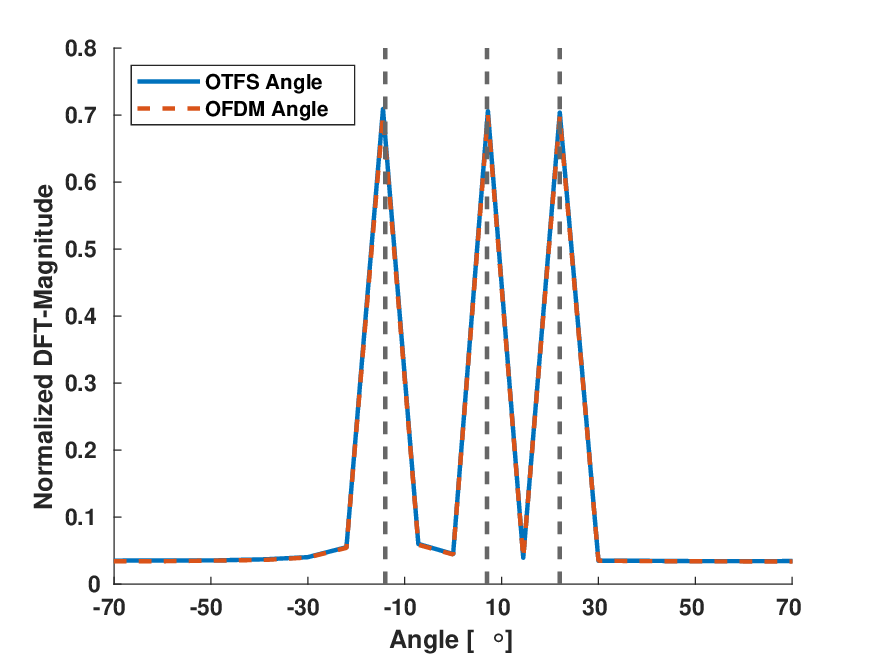}
        \caption{}
        \label{fig:angle_8x16}
    \end{subfigure}
    ~%
    \begin{subfigure}[t]{0.45\columnwidth}
        \centering
        \includegraphics[width=\columnwidth]{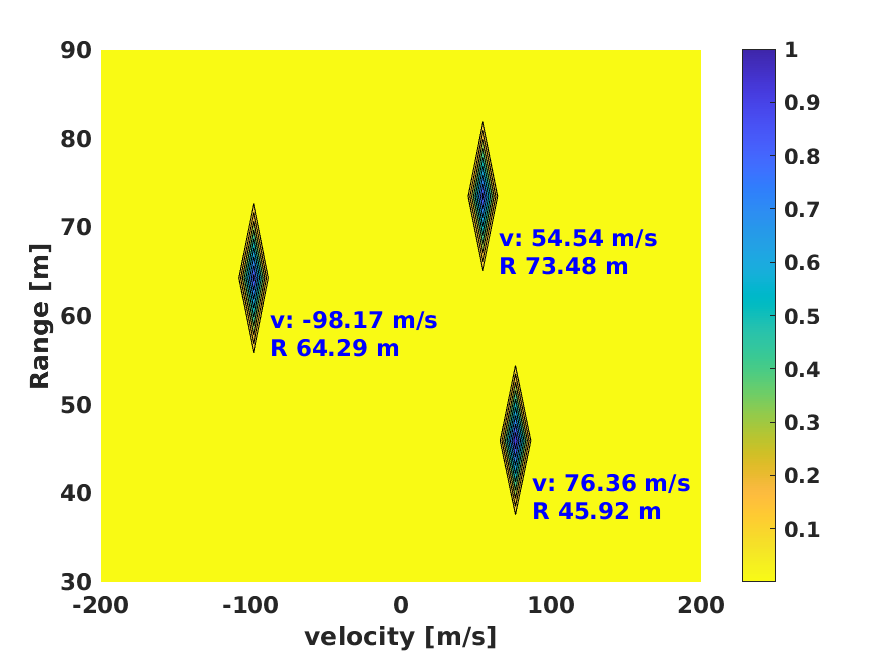}
        \caption{}
        \label{fig:2D_128x32}
    \end{subfigure}
    \begin{subfigure}[t]{0.45\columnwidth}
        \centering
        \includegraphics[width=\columnwidth]{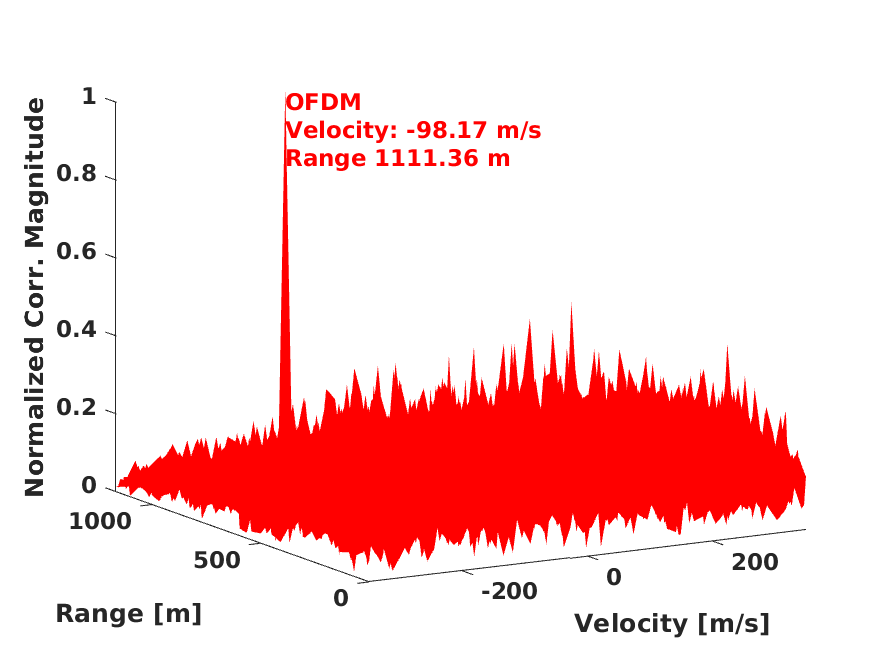}
        \caption{}
        \label{fig:Cor_OFDM_2D}
    \end{subfigure}
    ~%
    \begin{subfigure}[t]{0.45\columnwidth}
        \centering
        \includegraphics[width=\columnwidth]{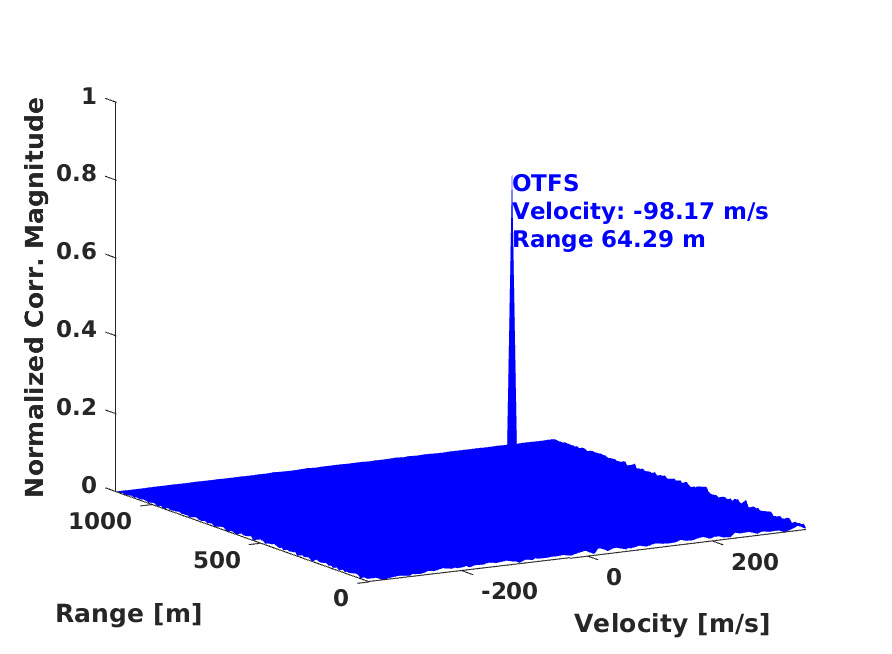}
        \caption{}
        \label{fig:Cor_OTFS_2D}
    \end{subfigure}
    \caption{
    (a) Angle estimation with ground truth indicated by black dashed lines. (b) 2D Cross-correlation based range and velocity estimation (the ground truth is shown on the figure).   Range and velocity estimation of the second target with (c) OFDM waveforms and (d) OTFS waveforms.
    }
\end{figure}

We consider a system with \(N_t = 4\) transmit antennas and \(N_r = 16\) receive antennas.
The parameters of the targets/paths are listed in~\cref{tab:share_parameters}; under these conditions, all targets are well separated.
To estimate the target angles, we employ the DFT-based approach described in~\cref{sec:shared_angle}, and the resulting estimates are illustrated in~\cref{fig:angle_8x16}.
As angle estimation depends primarily on spatial frequency, both OTFS and OFDM waveforms correctly identify all three targets.
Moreover, by averaging across all bins for OTFS and across all subcarriers for OFDM~\cite{xu2023abandwidth}, the resulting angle estimates appear smooth and well-defined.
The range and velocity are estimated using the 2D cross-correlation technique of~\cref{sec:shared_2D} and are shown in~\cref{fig:2D_128x32}.
For comparison with OFDM, we also include the OFDM approach of~\cite{xu2023abandwidth} (\cref{fig:Cor_OFDM_2D}) along with the proposed OTFS approach (\cref{fig:Cor_OTFS_2D}), corresponding to the second target.
While OTFS accurately recovers the target’s range and velocity, OFDM fails to estimate the range, mainly due to loss of orthogonality in this high Doppler scenario.

\subsubsection{DFT-Based Angle Estimation Performance}
\begin{figure}
    \centering
        \includegraphics[width=0.75\columnwidth]{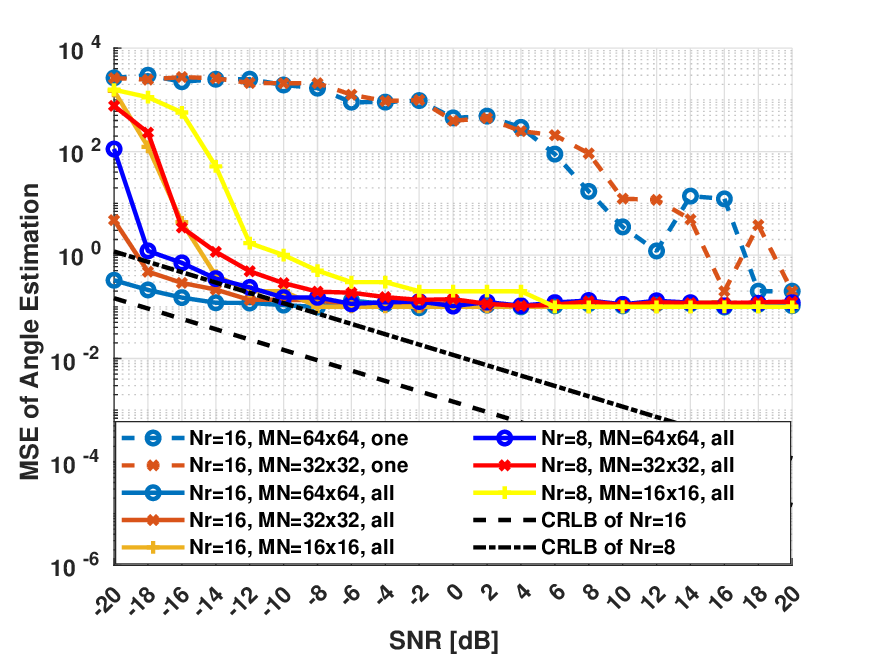}
    \caption{MSE and CRLB of coarse angle estimates. The estimation on a single bin is denoted as \textit{one}, and the estimation by averaging across all bins is denoted as \textit{all}.}
    \label{fig:angle_CRLB}
\end{figure}

To further evaluate the proposed algorithm, we perform a Monte Carlo simulation with $500$ trials.
In each trial of angle estimation, one target is randomly set with an angle in $[-60,60]^{\circ}$, while the remaining parameters are chosen from~\cref{tab:share_parameters}.
We obtain results for $N_r=8$ and $N_r=16$ and different OTFS grid sizes, \ie $M\times N=64\times 64$, $M\times N=32\times 32$ and $M\times N=8\times 8$.
\cref{fig:angle_CRLB} shows the MSE of the angle estimates obtained based on a single DD bin and also angle estimates obtained by averaging over all \(NM\) bins.
The corresponding CRLB, given by~\cref{eq:crlb_angle}, is also plotted for comparison.

We observe that averaging estimates across all bins significantly reduces the MSE and leads to faster convergence to the CRLB.
When the grid is sufficiently large, our results match the CRLB, even at low SNRs.
We also observe that increasing the value of $N_r$ results in lower MSE, also aligned with the derived angle CRLB.
In every configuration, the achieved MSE is constrained by the angle aperture, preventing the estimate from achieving the CRLB.

\subsubsection{Cross-Correlation-Based Range and Velocity Estimation Performance}
\textcolor{black}{We conducted Monte Carlo simulations, considering a single target, with $M=2048$ and $B=245.76$MHz as a common choice in modern wideband communication systems and $N=32$ and $T=0.2667$ms to ensure low latency communication.}
The
single target's range is randomly selected in $[50:0.5:100]\rm{m}$.
The target's velocity is randomly selected in $[-100:2:100]\rm{m/s}$.
All other parameters are chosen from~\cref{tab:share_parameters}.
In this case, 
the range resolution is $R_{\rm res}=c/(2M\Delta f)=0.61\rm{m}$ and the target is almost on the delay grid.
The
velocity resolution is $v_{res}=\lambda/(2N\Delta t)=23.09\rm{m/s}$ and the target is usually off the Doppler grid, so the target suffers from fractional Doppler.

\begin{figure}
        \centering
        \includegraphics[width=0.75\columnwidth]{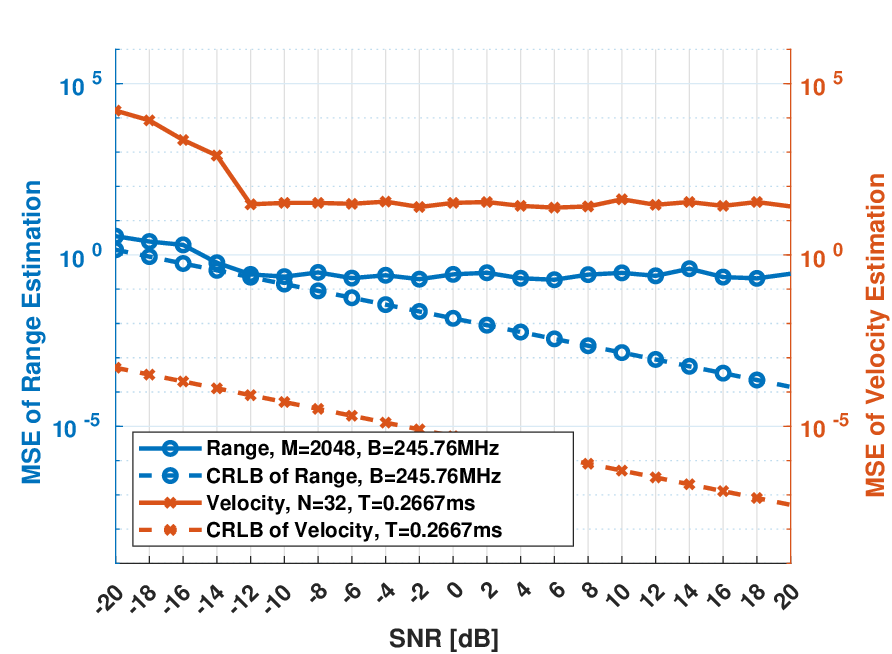}
        \caption{MSE and CRLB of range and velocity estimation.}
        \label{fig:velocity_CRLB_frac}
\end{figure}

\cref{fig:velocity_CRLB_frac} displays the MSE of the range and velocity estimates obtained via the cross-correlation approach of~\cref{sec:shared_2D}.
The corresponding CRLBs, given by~\cref{eq:crlb_tau} after multiplication  by $(c/2)^2$, to convert the delay CRLB  to range CRLB, and of~\cref{eq:crlb_nu} after multiplication  by $(\lambda/2)^2$, to convert the Doppler CRLB to velocity CRLB, are also plotted for comparison.

It can be seen that the MSEs are constrained by the grid spacing, as in the case of angle estimation.
We also observe that the range estimate is able to approach its CRLB before its MSE is constrained by the grid spacing, while the velocity estimate is far away from its CRLB due to the fractional Doppler.



\begin{figure}
    \centering
    \includegraphics[width=0.75\columnwidth]{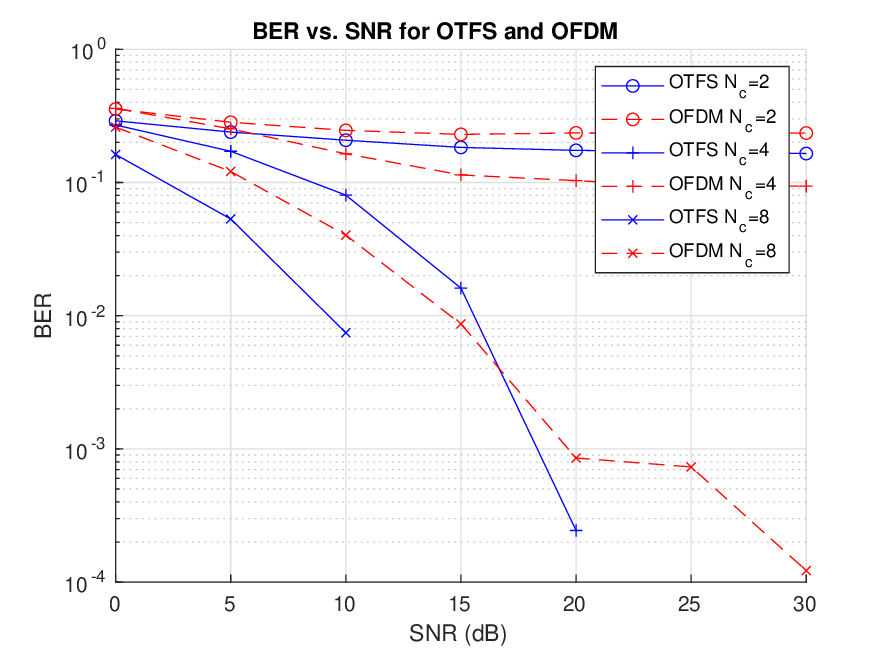}
    \caption{MIMO communication performance between the OTFS and OFDM waveforms under high Doppler channel. The system is equipped with $N_t=4$ and a different number of $N_c$.}
    \label{fig:BER_Rx}
\end{figure}

\subsubsection{Communication Performance}
The Bit Error Rate (BER) of the proposed DFRC system with the different number of receive antennas is shown in~\cref{fig:BER_Rx}.
As expected, in both OTFS and OFDM systems, the BER decreases as the number of receive antennas increases.
We can observe that, for the same number of receive antennas, the OTFS system has lower BER than the OFDM system.
The shared use of DD bins increases the data rate by \(N_t\), and adding more transmit antennas further boost it. However, larger \(N_t\) complicates equalization for MIMO symbol detection, and finding an effective algorithm for OTFS communication with large MIMO antennas remains an open challenge.

\subsection{Trading off Communication Rate  for Improved Sensing Using Private Bins and Virtual Array}\label{sec:VA_sim}

\begin{figure}
    \centering
    \begin{subfigure}[t]{0.45\columnwidth}
        \centering
        \includegraphics[width=\columnwidth]{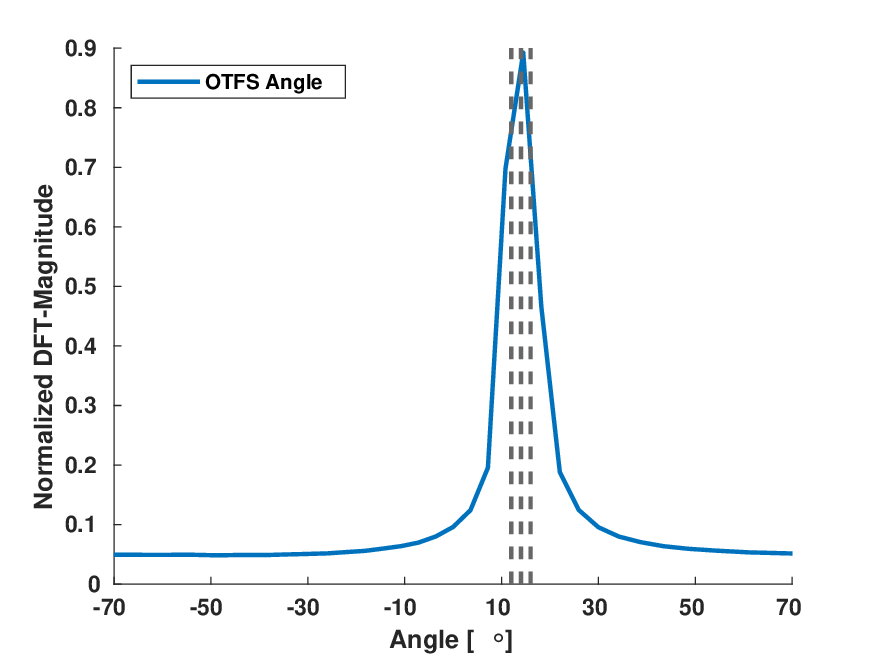}
        \caption{}
        \label{fig:OTFS_VA_DFT}
    \end{subfigure}
    ~%
    \begin{subfigure}[t]{0.45\columnwidth}
        \centering
        \includegraphics[width=\columnwidth]{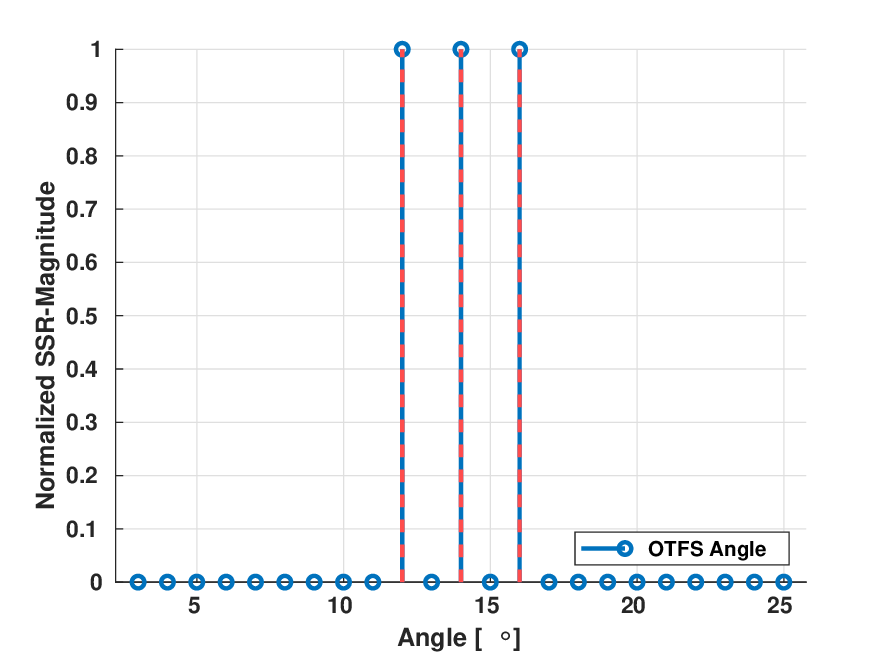}
        \caption{}
        \label{fig:OTFS_VA_SSR}
    \end{subfigure}
    \caption{While the DFT-based estimation captures one target only,  SSR-based angle estimation successfully captures all three targets. Dashed lines in (a) indicate the ground truth.}
\end{figure}

We take consider three targets at angles $[12, 14, 16]^{\circ}$,
$N_t=4$, $N_r=16$, and channel parameters as shown in~\cref{tab:share_parameters}.
In this case, with the targets located within $2^{\circ}$, \ie $\Delta\varphi=2^{\circ}$, only one peak is visible in the DFT, as shown in~\cref{fig:OTFS_VA_DFT}.
Next, we use the private bins designed as described in~\cref{sec:private}, to formulate and solve an SSR problem to refine angle estimation.

We use $N_p=4$ private bins in total. For each transmit antenna $i$, we left $\mathcal{E}_i=\set{[0,0],[1,1],[2,2]}$ empty and place symbols on the remaining DD  bins. In fact, any $3$ DD bins of each antenna could have chose to be left empty.
After performing the ISFFT, we obtain the TF domain signal where we introduce TF private bins $\mathcal{P}_0=[0,0]$, $\mathcal{P}_1=[1,1]$, $\mathcal{P}_2=[2,2]$, and $\mathcal{P}_3=[3,3]$, and enforce zeros at locations $\mathcal{Z}_0=\set{[1,1],[2,2],[3,3]}$, $\mathcal{Z}_1=\set{[0,0],[2,2],[3,3]}$, $\mathcal{Z}_2=\set{[0,0],[1,1],[3,3]}$, and $\mathcal{Z}_3=\set{[0,0],[1,1],[2,2]}$.

We then use the signal received on the private TF bins and also the \textit{coarse estimates} obtained from the method described in~\cref{sec:shared} to formulate the SSR problem as described in~\cref{sec:private}.

\textcolor{black}{
Each row of $\mathbf{\Phi}$ corresponds to a private TF bin on one receive antenna.
Each column of $\mathbf{\Phi}$ is in the form \textcolor{black}{of~\cref{eq:SSR_ratio}} corresponding to a grid point of the  angle-Doppler-delay space, discretized around \textit{coarse estimates} $(\varphi_j, \nu_j, \tau_j)$.
In this example, the angle space is discretized around $\varphi_j=14^{\circ}$
with $\delta \varphi=1^{\circ}$ and $W_{\varphi_j}=10^{\circ}$.
The Doppler space and delay space are around the parameters shown in the~\cref{tab:share_parameters} with $\delta\nu=0.1\Delta\nu$, $W_{\nu_j}=2\Delta\nu$, $\delta\tau=0.1\Delta\tau$, $W_{\tau_j}=2\Delta\tau$.
}%
The vector $\mathbf{r}$ {(see~\cref{overcomplete})} can be obtained by stacking signals on all TF private bins across all receive antennas.
Upon solving the SSR problem, the peaks align accurately with the ground truth (see ~\cref{fig:OTFS_VA_SSR}), which illustrates the ability of the virtual array to estimate multiple closely spaced targets where the physical array fails.


\begin{figure}
    \centering
    \includegraphics[width=0.75\columnwidth]{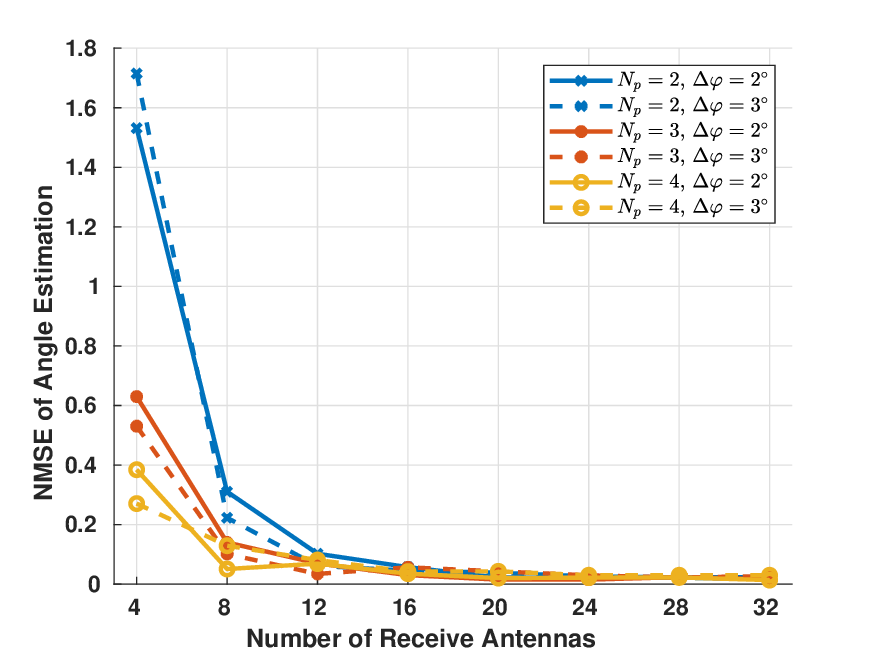}
    \caption{NMSE of SSR Angle estimation versus number of receive antennas,  for different numbers of private bins, and different target angle separation $\Delta\varphi$.  }
    \label{fig:VA_Angle_Err}
\end{figure}

\begin{figure}
\centering
\begin{subfigure}[t]{0.75\columnwidth}
    \centering
    \includegraphics[width=\columnwidth]{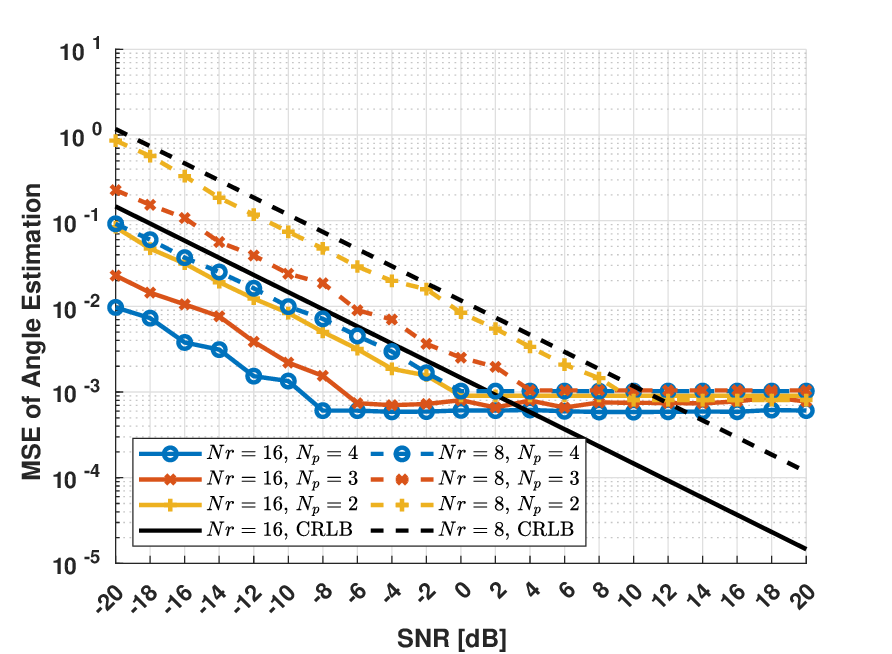}
    \caption{$\delta\varphi=0.1$}
    \label{fig:VA_Angle_Dis0.1}
\end{subfigure}
\begin{subfigure}[t]{0.75\columnwidth}
    \centering
    \includegraphics[width=\columnwidth]{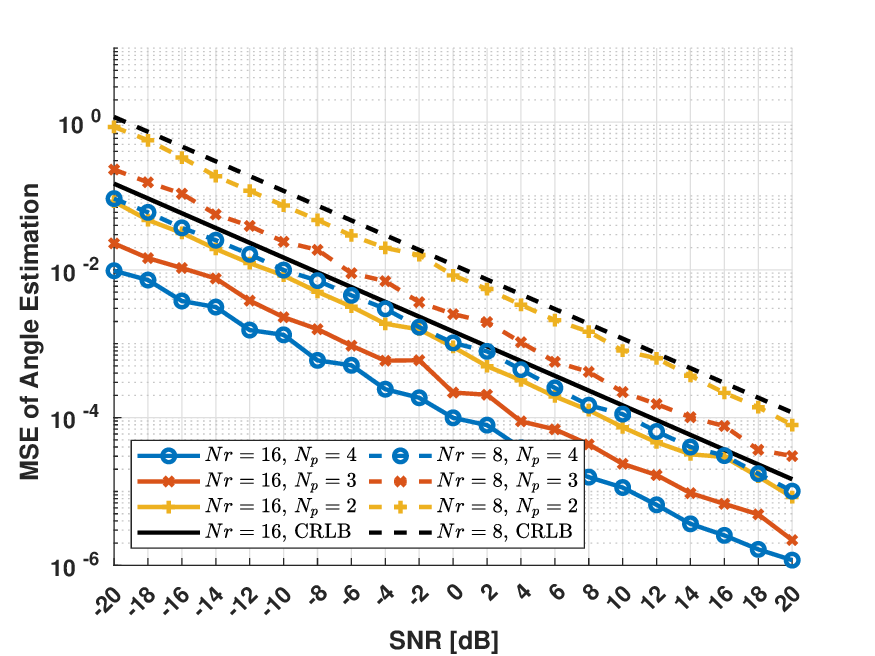}
    \caption{$\delta\varphi=0.01$}
    \label{fig:VA_Angle_Dis0.01}
\end{subfigure}
\caption{MSE and CRLB of SSR angle estimation.}
\end{figure}

\begin{figure}
    \centering
    \includegraphics[width=0.75\columnwidth]{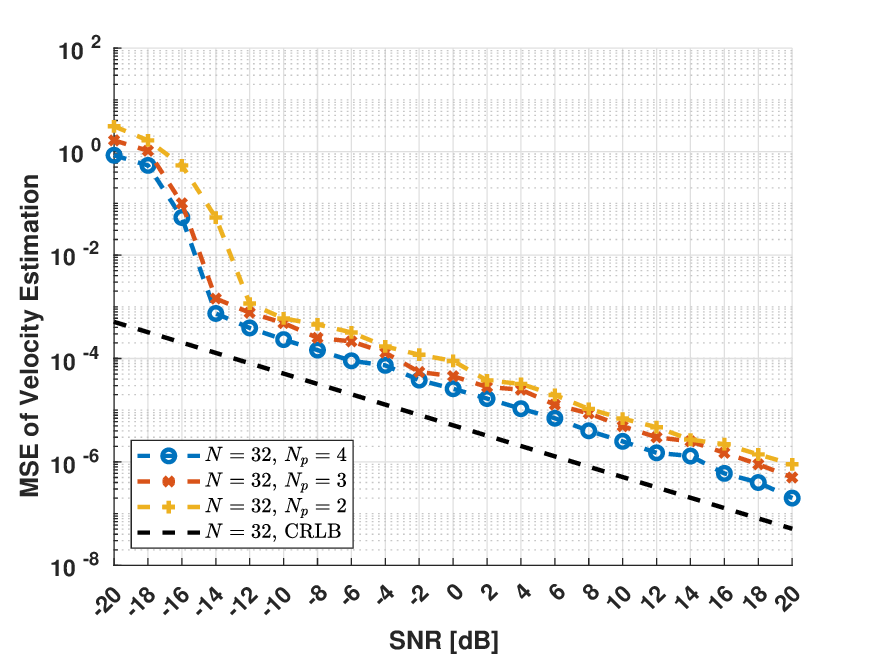}
    \caption{MSE and CRLB of SSR velocity estimation.}
    \label{fig:VA_velocity_CRLB}
\end{figure}

\subsubsection{SSR Angle Estimation Performance}

We examine how the number of private bins ($N_p$) and the target angle separation ($\Delta\varphi$) affect the detection performance of the SSR approach.
We conduct $500$ Monte Carlo runs without variance reduction, where in each run, we consider $3$ targets with angle in $[-60, 60]^{\circ}$ and other parameters are as shown in~\cref{tab:share_parameters}.
Various numbers of receive antennas, private bins, and $\Delta\varphi=2^{\circ}$ or $\Delta\varphi=3^{\circ}$ are considered.
In each run, the angle space is discretized around its \textit{coarse estimates} $\varphi_j$ with $\delta \varphi=1^{\circ}$ and $W_{\varphi_j}=10^{\circ}$.
The Doppler space and delay space are around the parameters shown in the~\cref{tab:share_parameters} with $\delta\nu=0.1\Delta\nu$, $W_{\nu_j}=2\Delta\nu$, $\delta\tau=0.1\Delta\tau$, $W_{\tau_j}=2\Delta\tau$.
The normalized mean squared error (NMSE) of angle estimates is depicted in~\cref{fig:VA_Angle_Err}.
We observe that any number of private bins results in low MSE when the number of receive antennas $N_r$ is large.
However, when $N_r$ is small, each additional private bin yields a considerable reduction in estimation error.
This result highlights the value of a virtual array for radar sensing with a limited number of antennas.

We then conduct $500$ Monte Carlo runs with variance reduction \textcolor{black}{(see~\cref{sec:private})}, to evaluate the effect of angle discretization ($\delta\varphi$) under a different number of private bins ($N_p$).
In each run, we consider $3$ targets in $[-60, 60]^{\circ}$ with $\Delta\varphi=2^{\circ}$ and $N_r=16$.
The $\delta\varphi=0.1^{\circ}$ or $\delta\varphi=0.01^{\circ}$ while other parameters and discretization are the same as above.
As depicted in~\cref{fig:VA_Angle_Dis0.1}, after performing the variance reduction, the MSE follows the CRLB trend but is limited by the resolution corresponding to \(\delta\varphi = 0.1^{\circ}\), making it difficult to drop below \(10^{-3}\).
By refining the discretization to $\delta\varphi=0.01^{\circ}$, we achieve further MSE reduction (see~\cref{fig:VA_Angle_Dis0.01}), demonstrating that the virtual array can do better than the CRLB.
This observation is consistent with the behavior shown in~\cref{fig:VA_Angle_Err}.

\subsubsection{Communication with the Use of Private Bins}
\begin{figure}
    \centering
    \includegraphics[width=0.70\columnwidth]{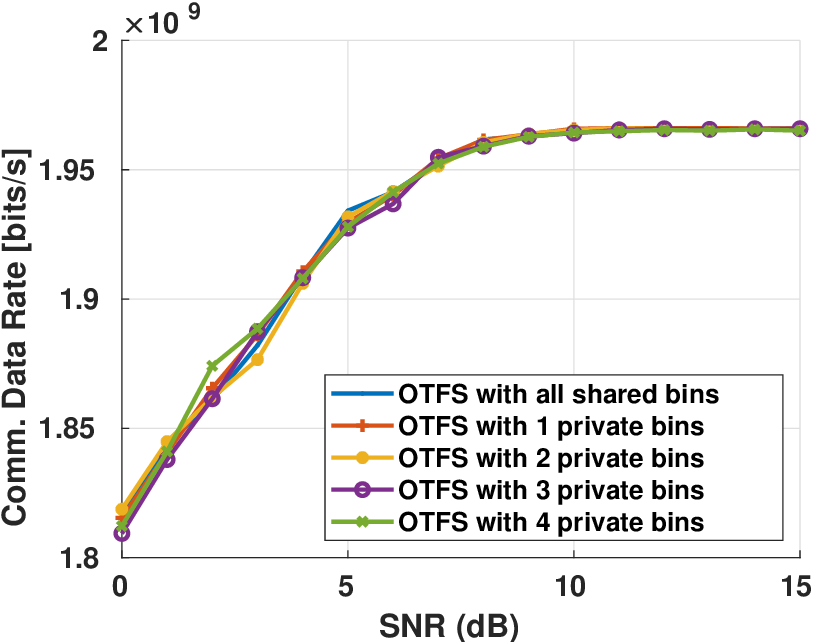}
    \caption{Communication Rate Loss Using Private Bins.}
    \label{fig:comm_rate}
\end{figure}

\textcolor{black}{
We evaluate the communication data rate of the proposed system with different numbers of private bins under different SNRs.
For the case of $N_t=4$, $N_c=8$, as shown in~\cref{fig:comm_rate}, the private bin design has similar communication performance compared to the all shared bin design.
}%
The loss of communication rate due to the use of private bins is rather small.
With $N_t=4$ and using QPSK symbols, each private bin leads to $3\times \log_2(4)\times 120\mathrm{kHz}=7.2\times 10^5$ [bits/s] communication loss.
When $4$ private bins are used, the percentage communication rate loss is $\frac{N_t\times 3}{N_t\times NM}=0.037\%$.

\subsection{SSR Fractional Doppler Estimation Performance}\label{sec:frac_performance}
Finally, we examine the virtual array for fractional Doppler estimation.
We conduct a Monte Carlo simulation with $500$ trials.
In each trial, a target is set with velocity from $[-100, 100]\rm{m/s}$, while the remaining parameters are chosen from~\cref{tab:share_parameters} and $N=32$.
For each SSR solver, the Doppler space is discretized around coarse estimates $\nu_j$ with $\delta\nu=0.02\Delta\nu$ and $W_{\nu_j}=3\Delta\nu$.
The angle space and delay space are around the parameters shown in the~\cref{tab:share_parameters} with $\delta \varphi=1^{\circ}$, $W_{\varphi_j}=10^{\circ}$, $\delta\tau=0.1\Delta\tau$, $W_{\tau_j}=2\Delta\tau$.
As can be seen from~\cref{fig:VA_velocity_CRLB}, the MSE is inflated by the noise when ${\rm SNR}<-12{\rm dB}$ and decreases as CRLB elsewhere.
This demonstrates the effectiveness of SSR in estimating the fractional Doppler.
Different from angle estimation, the use of private bins and SSR does not bring the MSE below CRLB in velocity estimation.
This is because the virtual array only increases the aperture of the spatial space and has no effect on the Doppler-delay resolution.
We also observe that more private bins can stabilize the estimation and slightly improve the MSE.
This is because more private bins will introduce more rows in $\mathbf{\Phi}$, helping mitigate the coherence between the matrix columns.

\section{Conclusion}\label{sec:conclusion}
We have proposed a novel DFRC MIMO OTFS system that is robust to Doppler shifts, can efficiently use the bandwidth for communication and sensing, and is equipped with a low-complexity, high-resolution radar parameter estimator.
DD domain bins are used in a shared fashion, while some TF bins are private to a small number of transmit antennas.
The shared bins enable high communication rates.
Introducing private bins enables the formation of a virtual array that improves the sensing performance at the cost of  some reduction of the DD domain symbols transmitted by each antenna, leading to a trade-
off between communication rate and sensing performance.
However, a small number of private bins, or equivalently, a very small amount of rate loss, suffices to achieve significant sensing gains.

\appendix[Derivation  of the CRLB]\label{apd:crlb}

\begin{table*}
\centering
\begin{minipage}{\textwidth}
\begin{align}
    \frac{\partial h_{n_r,k,l}}{\partial\tau_j}
    &= \frac{1}{NM} e^{j\pi n_r\sin(\varphi_j)} \beta_j e^{j\phi_j} \sum_{n=0}^{N-1}e^{-j2\pi(k-\nu_j N\Delta t)\frac{n}{N}}\sum_{m=0}^{M-1}e^{j2\pi \frac{m}{M}(l-\tau_j M\Delta f)}\bigl\{-j2\pi m\Delta f\bigr\}, \label{eq:derive_tau}\\
    \frac{\partial h_{n_r,k,l}}{\partial\nu_j}
    &= \frac{1}{NM} e^{j\pi n_r\sin(\varphi_j)} \beta_j e^{j\phi_j} \sum_{n=0}^{N-1}e^{-j2\pi(k-\nu_j N\Delta t)\frac{n}{N}}\bigl\{j2\pi n\Delta t\bigr\}\sum_{m=0}^{M-1}e^{j2\pi \frac{m}{M}(l-\tau_j M\Delta f)},  \label{eq:derive_nu}\\
    \frac{\partial h_{n_r,k,l}}{\partial \pi\sin(\varphi_j)}
    &= \frac{1}{NM} e^{j\pi n_r\sin(\varphi_j)} \bigl\{j n_r\bigr\} \beta_j e^{j\phi_j} \sum_{n=0}^{N-1}e^{-j2\pi(k-\nu_j N\Delta t)\frac{n}{N}}\sum_{m=0}^{M-1}e^{j2\pi \frac{m}{M}(l-\tau_j M\Delta f)}, \label{eq:derive_angle} \\
    \frac{\partial h_{n_r,k,l}}{\partial\phi_j}
    &= \frac{1}{NM} e^{j\pi n_r\sin(\varphi_j)} \beta_j e^{j\phi_j} \bigl\{j\bigr\} \sum_{n=0}^{N-1}e^{-j2\pi(k-\nu_j N\Delta t)\frac{n}{N}}\sum_{m=0}^{M-1}e^{j2\pi \frac{m}{M}(l-\tau_j M\Delta f)}. \label{eq:derive_phase}
\end{align}
\hrule
\end{minipage}
\end{table*}%

\begin{table*}
\centering
\begin{minipage}{\textwidth}
\begin{align}
    \mathbf{A}
    =\sum_{n_r=0}^{N_r-1}\sum_{k=0}^{N-1}\sum_{l=0}^{M-1}
    \begin{bmatrix}
        \parens*{\frac{\partial h_{n_r,k,l}}{\partial\tau_j}}\!^*\frac{\partial h_{n_r,k,l}}{\partial\tau_j} &
        \parens*{\frac{\partial h_{n_r,k,l}}{\partial\tau_j}}\!^*\frac{\partial h_{n_r,k,l}}{\partial\nu_j} &
        \parens*{\frac{\partial h_{n_r,k,l}}{\partial\tau_j}}\!^*\frac{\partial h_{n_r,k,l}}{\partial\pi\sin(\varphi_j)} &
        \parens*{\frac{\partial h_{n_r,k,l}}{\partial\tau_j}}\!^*\frac{\partial h_{n_r,k,l}}{\partial\phi_j} \\
        \parens*{\frac{\partial h_{n_r,k,l}}{\partial\nu_j}}\!^*\frac{\partial h_{n_r,k,l}}{\partial\tau_j} &
        \parens*{\frac{\partial h_{n_r,k,l}}{\partial\nu_j}}\!^*\frac{\partial h_{n_r,k,l}}{\partial\nu_j} &
        \parens*{\frac{\partial h_{n_r,k,l}}{\partial\nu_j}}\!^*\frac{\partial h_{n_r,k,l}}{\partial\pi\sin(\varphi_j)} &
        \parens*{\frac{\partial h_{n_r,k,l}}{\partial\nu_j}}\!^*\frac{\partial h_{n_r,k,l}}{\partial\phi_j} \\
        \parens*{\frac{\partial h_{n_r,k,l}}{\partial\pi\sin(\varphi_j)}}\!^*\frac{\partial h_{n_r,k,l}}{\partial\tau_j} &
        \parens*{\frac{\partial h_{n_r,k,l}}{\partial\pi\sin(\varphi_j)}}\!^*\frac{\partial h_{n_r,k,l}}{\partial\nu_j} &
        \parens*{\frac{\partial h_{n_r,k,l}}{\partial\pi\sin(\varphi_j)}}\!^*\frac{\partial h_{n_r,k,l}}{\partial\pi\sin(\varphi_j)} &
        \parens*{\frac{\partial h_{n_r,k,l}}{\partial\pi\sin(\varphi_j)}}\!^*\frac{\partial h_{n_r,k,l}}{\partial\phi_j} \\
        \parens*{\frac{\partial h_{n_r,k,l}}{\partial\phi_j}}\!^*\frac{\partial h_{n_r,k,l}}{\partial\tau_j} &
        \parens*{\frac{\partial h_{n_r,k,l}}{\partial\phi_j}}\!^*\frac{\partial h_{n_r,k,l}}{\partial\nu_j} &
        \parens*{\frac{\partial h_{n_r,k,l}}{\partial\phi_j}}\!^*\frac{\partial h_{n_r,k,l}}{\partial\pi\sin(\varphi_j)} &
        \parens*{\frac{\partial h_{n_r,k,l}}{\partial\phi_j}}\!^*\frac{\partial h_{n_r,k,l}}{\partial\phi_j} &
    \end{bmatrix}. \label{eq:derive_05}
\end{align}
\hrule
\end{minipage}
\end{table*}

\begin{table*}
\centering
\begin{minipage}{\textwidth}
\begin{align}
    \sum_{k=0}^{N-1}\abs*{\sum_{n=0}^{N-1}1\cdot e^{-j2\pi(k-\nu_j N\Delta t)\frac{n}{N}}}^2
    &= N\sum_{n=0}^{N-1}\abs*{1}^2 = N\cdot N. \label{eq:Parseval_1}\\
    \sum_{l=0}^{M-1}\abs*{\sum_{m=0}^{M-1}m\cdot e^{j2\pi \frac{m}{M}(l-\tau_j M\Delta f)}}^2
    &= M\sum_{m=0}^{M-1}\abs*{m}^2=M\cdot\frac{M(M-1)(2M-1)}{6}. \label{eq:Parseval_2}\\
    \sum_{n_r=0}^{N_r-1}\sum_{k=0}^{N-1}\sum_{l=0}^{M-1}\parens*{\frac{\partial h_{n_r,k,l}}{\partial\tau_j}}\!^*\frac{\partial h_{n_r,k,l}}{\partial\tau_j}
    &= \abs*{\frac{1}{NM}}^2\abs*{e^{j\pi n_r\sin(\varphi_j)}}^2 \abs*{\beta_j}^2 \abs{e^{j\phi_j}}^2 \sum_{n_r=0}^{N_r-1} 1 \notag\\
    &\times \sum_{k=0}^{N-1}\abs*{\sum_{n=0}^{N-1}1\cdot e^{-j2\pi(k-\nu_j N\Delta t)\frac{n}{N}}}^2
    (2\pi \Delta f)^2\sum_{l=0}^{M-1}\abs*{\sum_{m=0}^{M-1}m\cdot e^{j2\pi \frac{m}{M}(l-\tau_j M\Delta f)}}^2 \notag\\
    &= \abs{\beta_j}^2 N_r 4\pi^2(\Delta f)^2 \frac{(M-1)(2M-1)}{6}. \label{eq:ease} \\
    \sum_{n_r=0}^{N_r-1}\sum_{k=0}^{N-1}\sum_{l=0}^{M-1}\parens*{\frac{\partial h_{n_r,k,l}}{\partial\pi\sin(\varphi_j)}}\!^*\frac{\partial h_{n_r,k,l}}{\partial\pi\sin(\varphi_j)}
    &= \abs*{\frac{1}{NM}}^2\abs*{e^{j\pi n_r\sin(\varphi_j)}}^2 \abs*{\beta_j}^2 \abs{e^{j\phi_j}}^2
    \sum_{n_r=0}^{N_r-1} n_r^2
    \notag\\
    &\times \sum_{k=0}^{N-1}\abs*{\sum_{n=0}^{N-1}1\cdot e^{-j2\pi(k-\nu_j N\Delta t)\frac{n}{N}}}^2
    \sum_{l=0}^{M-1}\abs*{\sum_{m=0}^{M-1}1\cdot e^{j2\pi \frac{m}{M}(l-\tau_j M\Delta f)}}^2 \notag\\
    &= \abs{\beta_j}^2 \frac{N_r(N_r-1)(2N_r-1)}{6}. \label{eq:ease_angle}
\end{align}
\hrule
\end{minipage}
\end{table*}


\begin{table*}
\centering
\begin{minipage}{\textwidth}
\begin{align}
    \mathbf{I}_{\bftt_j \bftt_j}
    =
    \frac{2\abs{\beta_j}^2P_{\rm avg}}{N_0} N_r\cdot
    \underbrace{
    \begin{bmatrix}
        4\pi^2 (\Delta f)^2 \frac{(M-1)(2M-1)}{6} &
        -\pi^2 (N-1)(M-1) &
        -\pi \frac{N_r-1}{2}\Delta f (M-1) &
        -\pi \Delta f (M-1) \\
        -\pi^2 (N-1)(M-1) &
        4\pi^2 (\Delta t)^2 \frac{(N-1)(2N-1)}{6} &
        \pi \frac{N_r-1}{2}\Delta t (N-1) &
        \pi \Delta t (N-1) \\
        -\pi \frac{N_r-1}{2}\Delta f (M-1) &
        \pi \frac{N_r-1}{2}\Delta t (N-1) &
        \frac{(N_r-1)(2N_r-1)}{6} &
        \frac{N_r-1}{2} \\
        -\pi \Delta f (M-1) &
        \pi \Delta t (N-1) &
        \frac{N_r-1}{2} &
        1
    \end{bmatrix}
    }_{\mathbf{C}}. \label{eq:derive_3}
\end{align}
\hrule
\end{minipage}
\end{table*}


Based on $\frac{\partial h_{n_r,k,l}}{\partial\bftt_j}$, which are shown in~\cref{eq:derive_tau,eq:derive_nu,eq:derive_angle,eq:derive_phase},
we can write~\cref{eq:FIM_6} as
\begin{align}
    \mathbf{I}_{\bftt_j \bftt_j}
    &= \frac{2P_{\rm avg}}{N_0} \Re\{\mathbf{A}\},  
\end{align}
where $\mathbf{A}$ is shown in~\cref{eq:derive_05}.
\textcolor{black}{In order to obtain closed-form expressions for easier interpretation, we consider the case in which $N$ and $M$ are large enough, in which case the targets are on the grid, \ie, \(\nu_j N\Delta t\) and \(\tau_j M\Delta f\) are integers, and $N_r$ is large. In that asymptotic case,}
$\mathbf{I}_{\bftt_j \bftt_j}$ is as shown in
\cref{eq:derive_3}.
\textcolor{black}{
In obtaining~\cref{eq:derive_3} we used the equalities of shown in~\cref{eq:Parseval_1,eq:Parseval_2,eq:ease,eq:ease_angle},
which are based on the Discrete Parseval's Theorem and expressions for the finite sum.
}
Denote $\mathbf{C}$ as a $2\times2$ block matrix
\begin{align}
    \mathbf{C}=
    \begin{bmatrix}
        \mathbf{D}_0 & \mathbf{E} \\
        \mathbf{E}\trn & \mathbf{D}_1
    \end{bmatrix}.
\end{align}
We employ the Woodbury matrix identity
{\small
\begin{align*}
\mathbf{C}^{-1}
&=
\begin{bmatrix}
    (\mathbf{D}_0-\mathbf{E}\mathbf{D}_1^{-1}\mathbf{E}\trn)^{-1} & 0 \\
    0 & (\mathbf{D}_1-\mathbf{E}\trn\mathbf{D}_0^{-1}\mathbf{E})^{-1}
\end{bmatrix}
\begin{bmatrix}
    \mathbf{I} & \ldots \\
    \ldots & \mathbf{I}
\end{bmatrix} 
\end{align*}
}%
and
{\footnotesize
\begin{align}
    &\mathbf{D}_1^{-1}=
    \begin{bmatrix}
        \frac{12}{(N_r^2-1)} & -\frac{6}{N_r+1} \\
        -\frac{6}{N_r+1} & \frac{4N_r-2}{N_r+1}
    \end{bmatrix}, \label{eq:d1_inv}\\
    &\mathbf{E}\mathbf{D}_1^{-1}\mathbf{E}\trn=
    \begin{bmatrix}
        \pi^2(\Delta f)^2(M-1)^2 & -\pi^2(N-1)(M-1) \\
        -\pi^2(N-1)(M-1) & \pi^2(\Delta t)^2(N-1)^2
    \end{bmatrix}, \\
    &\mathbf{D}_0-\mathbf{E}\mathbf{D}_1^{-1}\mathbf{E}\trn=
    \begin{bmatrix}
        \frac{\pi^2(\Delta f)^2(M^2-1)}{3} & 0 \\
        0 & \frac{\pi^2(\Delta t)^2(N^2-1)}{3}
    \end{bmatrix}. \label{eq:d0_inv}
\end{align}
}%
Then we can obtain~\cref{eq:crlb_tau,eq:crlb_nu} by
\begin{align*}
    \begin{bmatrix}
        \tau_j^{\rm CRLB} & \ldots \\
        \ldots & \nu_j^{\rm CRLB}
    \end{bmatrix}
    = \frac{N_0}{2\abs{\beta_j}^2P_{\rm avg}}\frac{1}{N_r}\cdot(\mathbf{D}_0-\mathbf{E}\mathbf{D}_1^{-1}\mathbf{E}\trn)^{-1}
\end{align*}
Similarly, we can get~\cref{eq:crlb_angle}.

\bibliographystyle{IEEEtran}
\bibliography{bib,ref}


\end{document}